\documentclass[lettersize,journal]{IEEEtran}
\usepackage{amsmath,amsfonts}
\usepackage{algorithmic}
\usepackage{algorithm}
\usepackage{array}
\usepackage[caption=false,font=normalsize,labelfont=sf,textfont=sf]{subfig}
\usepackage{gensymb}
\usepackage{stfloats}
\usepackage{url}
\usepackage{verbatim}
\usepackage{graphicx}
\usepackage{cite}
\usepackage{multirow}
\usepackage{diagbox}
\usepackage{hyperref}
\usepackage[colorinlistoftodos]{todonotes}

\newcommand\Qunit{$\mathrm{m}^{3}.\mathrm{s}^{-1}$}

\newcommand\reana{\textcolor{blue}{\mbox{ }}}
\newcommand\fct{\textcolor{red}{\mbox{ }}}

\newcommand\sophie{\textcolor{black}}

\begin{document}

\title{Remote Sensing Data Assimilation with a Chained Hydrologic-hydraulic Model for Flood Forecasting}

\author{IEEE Publication Technology,~\IEEEmembership{Staff,~IEEE,}
\thanks{This paper was produced by the IEEE Publication Technology Group. They are in Piscataway, NJ.}
\thanks{Manuscript received April 19, 2021; revised August 16, 2021.}}

\author{
        Thanh~Huy~Nguyen,
        Andrea~Piacentini,
        Sophie~Ricci,
        Ludovic~Cassan,
        Simon~Munier,
        Quentin~Bonassies,
        and Raquel~Rodriguez-Suquet
\thanks{This work was supported in part by the Centre National d'Études Spatiales (CNES) and in part by the Centre Européen de Recherche et de Formation Avancée en Calcul Scientifique (CERFACS) within the framework of the Space for Climate Observatory (SCO). (Corresponding author: Thanh Huy Nguyen.)}
\thanks{Thanh Huy Nguyen was with the CERFACS, 31057 Toulouse Cedex 1, France, and also with the CECI Laboratory, CERFACS/CNRS UMR 5318, 31057 Toulouse Cedex 1, France. He is now with the Environmental Research and Innovation department, Luxembourg Institute of Science and Technology (LIST), L-4326 Esch-sur-Alzette, Luxembourg  (e-mail: thanh-huy.nguyen@list.lu).}
\thanks{Sophie Ricci, Ludovic Cassan and Quentin Bonassies are with the CERFACS, 31057 Toulouse Cedex 1, France, and also with the CECI Laboratory, CERFACS/CNRS UMR 5318, 31057 Toulouse Cedex 1, France (e-mail: ricci@cerfacs.fr, cassan@cerfacs.fr, bonassies@cerfacs.fr).}
\thanks{Andrea Piacentini is with the CERFACS, 31057 Toulouse Cedex 1, France
(piacentini.palm@gmail.com).}
\thanks{Simon Munier is with the Centre National de Recherches Météorologiques (CNRM), UMR 3589, 31057 Toulouse Cedex 1, France
(simon.munier@meteo.fr).}
\thanks{Raquel Rodriquez Suquet is with the Centre National d'Études Spatiales (CNES), 31401 Toulouse Cedex 9, France (e-mail: raquel.rodriguezsuquet@cnes.fr).}
}



\maketitle

\begin{abstract}
A chained hydrologic-hydraulic model is implemented using predicted runoff from a large-scale hydrologic model (namely ISBA-CTRIP) as inputs to local and high-fidelity hydrodynamic models (TELEMAC-2D) in order to issue forecast of water level and flood extent. The uncertainties in the hydrological forcing and in friction parameters are reduced by an Ensemble Kalman Filter that jointly assimilates in-situ water level measurements and flood extent maps derived from satellite Earth observations. The data assimilation framework is cycled in a real-time forecasting configuration. A cycle consists of a reanalysis phase and a forecast phase. Over the reanalysis, observations up to the present are assimilated. An ensemble is then initialized from the last analyzed states and issued forecasts for the next 36~hr. Three strategies of forcing data for this forecast are investigated: (i) using CTRIP predicted runoff for reanalysis and forecast, (ii) using observed discharge for reanalysis and then CTRIP runoff for forecast and (iii) using observed discharge for reanalysis and keep a persistent discharge value for forecast. It was shown that the data assimilation strategy provides a reliable reanalysis of past flood events in the riverbed and in the floodplain. The combination of observed forcing for the reanalysis and CTRIP-predicted forcing for the forecast provides the most accurate water levels and flood extent results. For all three strategies, the quality of the forecast decreases as the lead time increases, especially beyond 18~hr. When the errors in the CTRIP forcing are non-stationary, the forecast capability may be reduced. This work demonstrates that the forcing provided by a hydrologic model, while imperfect, can be efficiently used as input to a local hydraulic model to issue reanalysis and forecasts, thanks to the assimilation of in-situ and remote sensing observations. 
\end{abstract}

\begin{IEEEkeywords}
Hydrodynamic model, Large-scale hydrology, Forecast, Data assimilation, Remote sensing, Flood extent.
\end{IEEEkeywords}

\section{Introduction}
\IEEEPARstart{E}{arly} warning and prediction of flood events have become increasingly more important as the occurrence and intensity of floods have increased in recent decades, especially in the context of climate change \cite{EMDAT2021}. For operational flood forecasting, the challenges lie in producing reliable forecasts under the constraints of computational resources and processing time required for operational forecasting. Hydrodynamic or hydraulic models---which are fundamental to estimating water levels and flood extent---require observed data from stream gauge networks. However, securing this data can be challenging due to either a lack of in-situ gauge data or the unreliability of measurements during intense flood events. As a result, prescribing water level and/or discharge as forcing time-series at the upstream and lateral boundary conditions (BCs) of the models can become problematic. However, such information is not always available, either due to a lack of in-situ gauging data, or due to the unreliability of measurements taken during intense flood events. 
Furthermore, constraining the hydraulic model to a liquid BC derived solely from observed forcing time-series limits its forecast lead time within the transfer time of the river network. To extend forecast lead times, it is necessary to introduce predicted inflow discharges into the hydrodynamic model, such as those simulated by large-scale hydrologic models. However, due to their coarse resolution and simplified representation of physical processes, these large-scale hydrologic models introduce uncertainty into the forcing of hydrodynamic models, leading to inaccuracies in flood forecasts. Improvements in large-scale hydrologic outputs are therefore imperative; this could be done through the assimilation of hydraulic variables into hydrodynamic models \cite{hostache2018near,DiMauro2021,grimaldi2016remote}. 

Flood simulation and forecasting capabilities have been greatly improved by advances in data assimilation (DA). DA combines observations as they become available with numerical models to reduce uncertainties in the model state, parameters, and/or BCs. 
A classical DA approach is the assimilation of water surface elevation data, either from in-situ gauge measurements, from altimetry satellites, or retrieved from remotely sensed RS images using flood edge location information combined with complementary Digital Elevation Model data \cite{grimaldi2016remote,garambois2020variational, Dasgupta2021review}. Satellite synthetic aperture radar (SAR) data are particularly advantageous for flood studies because they provide global, all-weather, day-and-night coverage of the continental water surface, which is characterized by low backscatter values resulting from the specular reflection of the incident radar pulses \cite{martinis2015flood}. Many research works have proposed the assimilation of RS-derived water surface elevation, as summarized in \cite{revilla2016integrating}. The need to retrieve water surface elevation from flood extents can be avoided with a direct assimilation of SAR-derived flood probability maps \cite{hostache2018near} or even flood extent maps \cite{nguyenagu2022}.

In the present paper, the 2D flood extent observations derived from remote sensing (RS) images, namely Sentinel-1 SAR images, are jointly assimilated with in-situ water level observations using an Ensemble Kalman Filter (EnKF). The DA algorithm reduces the uncertainties in the hydrodynamic model parameters as well as in the forcing inputs provided by in-situ measurement at gauging stations or by large-scale hydrologic models. 
We propose a chained hydrologic-hydrodynamic workflow dedicated to real-time forecasting of hydraulic variables over a reach of the Garonne River. This is orchestrated as a ensemble-based cycled DA procedure that stands in an analysis step over an assimilation window up to the current time, followed by a forecast step from the current time to the expected lead time. Different uses of observed or simulated forcing time-series, over the reanalysis and forecast phases, are presented for an overflowing event in early 2021. These strategies are compared based on the accuracy of the predicted water levels and flood extents, for lead time up to +36~hr.

The remainder of the article is organized as follows. Section~\ref{modelsdata} describes the hydrologic and hydrodynamic models used in this work, while Section~\ref{DA} presents the DA workflow for reanalysis and forecast. Section~\ref{results} provides the results with comprehensive assessments in the control and observation spaces, for the reanalysis experiments and the forecast at increasing lead-times. Conclusion and perspectives are lastly provided in Section~\ref{concpersp}.

\begin{figure}[!t]
\centering
\includegraphics[width=0.5\textwidth]{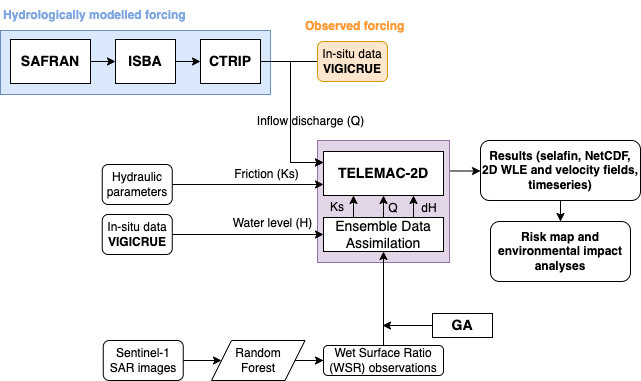}
\caption{ Workflow of the chained hydrologic-hydraulic model and the proposed DA strategy.}
\label{fig_1}
\end{figure}

\section{Models And Data}
\label{modelsdata}
\subsection{Hydrologic and hydrodynamics models}

\begin{figure}[!t]
\centering
\includegraphics[trim=0.6cm 0.6cm 0.6cm 0.6cm,clip, width=0.5\textwidth]{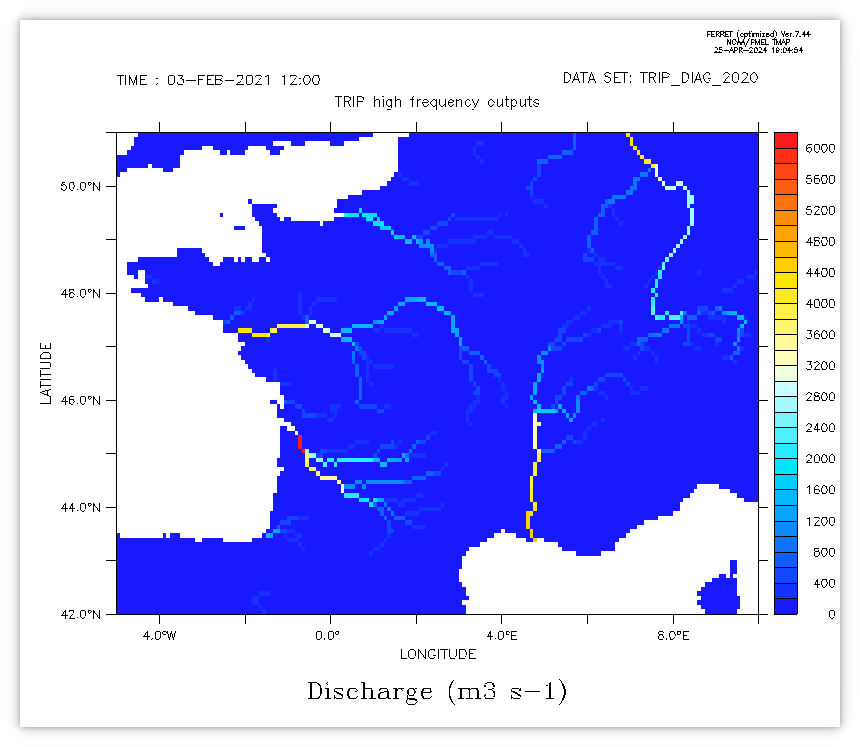}
\caption{Discharge (\Qunit) simulated by CTRIP over France and neighboring countries on 03/02/2021 12:00.}
\label{fig_2}
\end{figure}

In the present work, the BC for a local-scale and high-fidelity hydrodynamic model with TELEMAC-2D\footnote{\url{www.opentelemac.org}}  (T2D) is provided by the large-scale hydrologic model ISBA-CTRIP \cite{essd-14-2239-2022,decharme2019recent}. 
The general workflow for this strategy is shown in Figure~\ref{fig_1}. As mentioned above, the chained multi-physics and multi-scale modeling approach involves a direct supply of runoff forcing data to the T2D hydraulic model by a hydrologic model (blue block in Figure~\ref{fig_1}), as an alternative to  observed streamflow from an in-situ gauging station (orange block in Figure~\ref{fig_1}). The ISBA-CTRIP hydrologic model results from coupling the ISBA\footnote{Interaction Sol-Biosph\`{e}re-Atmosph\`{e}re} land surface model (LSM) \cite{noilhan1989simple} and a CNRM-modified version of the TRIP\footnote{Total Runoff Integrating Pathways} river routing model (RRM) \cite{oki1998design}. 
This present work is based on the recent CTRIP version from \cite{decharme2019recent,decharme2012global} that provides a gridded map of discharge, illustrated in Figure~\ref{fig_2} over France on 03/02/2021.
The CTRIP simulation time step is set to 1~hr, and the output streamflow time step is also hourly. LSMs simulate the energy and water balance at the soil-atmosphere-vegetation interface, while RRMs simulate the lateral transfer of freshwater towards the continent-ocean interface. The ISBA model is defined at the global scale on a $0.5^\circ \times 0.5^\circ$ regular grid, which establishes the energy and water budget over continental surfaces, considering a three-layer soil. It provides a diagnosis of surface runoff and gravitational drainage, which are later used as forcing inputs for CTRIP. The CTRIP model is defined on a regular latitude-longitude grid at the resolution of $1/12^\circ$ and follows a river network to transfer water laterally from one cell to another, down to the interface with the ocean. The uncertainty in these simulated discharges mainly stems from uncertainties in the LSM inputs (i.e. precipitation), RRM parameters and catchment description.

\subsection{Study area}
\begin{figure*}[!t]
\centering
\includegraphics[width=0.75\textwidth]{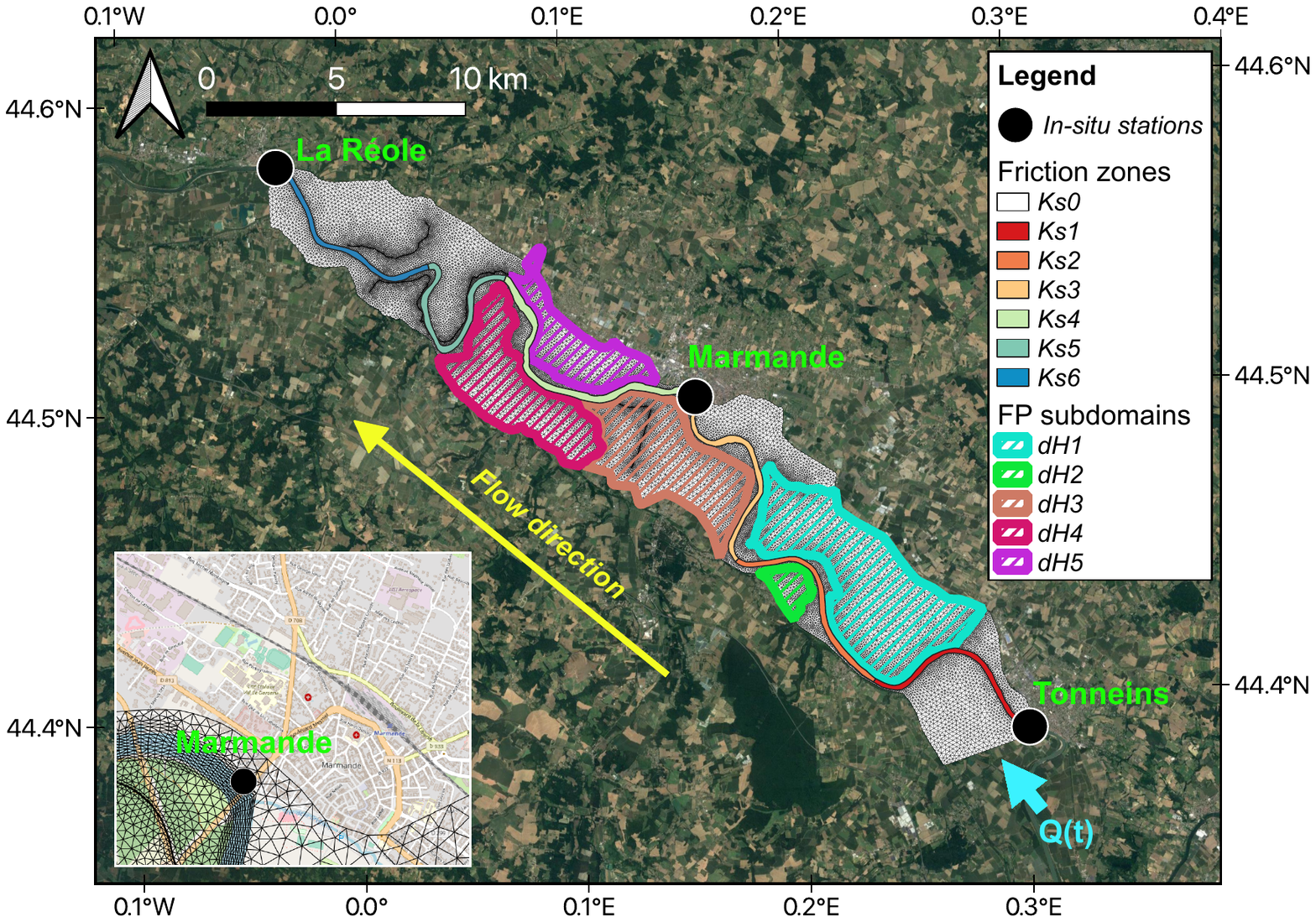}
\caption{TELEMAC-2D Garonne Marmandaise domain (from \cite{nguyenagu2022}). The in-situ observing stations are indicated as black circles. The different river friction zones are indicated as colored segments of the Garonne River. The floodplain is divided into five subdomains that are hatched in different colors. The inset figure shows the urban area of Marmande where its namesake gauge station is located.}
\label{fig_3}
\end{figure*}

The study area is shown in Figure~\ref{fig_3}. It extends over a 50-km stretch of the Garonne River between Tonneins and La Réole (rectangle in Figure~\ref{fig_3}). Gauging stations operated by EauFrance and used for the VigiCrue platform, located at Tonneins, Marmande and La Réole (black circles in Figure~\ref{fig_3}), provide water level measurements every 5 to 15 minutes. The local T2D hydraulic model over this reach was developed by EDF and presented in \cite{besnard2011comparaison,NguyenTGRS2022}. The Strickler friction coefficient \cite{gauckler1867etudes} is assumed to be uniform over each of six segments of the riverbed $K_{s_k}$ (with $k \in [1, 6]$), indicated by solid-colored segments in Figure~\ref{fig_3}, and also uniform over the entire floodplain ($K_{s_0}$). In addition, the limited number of in-situ gauging observations introduces errors in the formulation of the rating curve used to translate the observed water level into discharge, especially for high flows that would require extrapolation beyond the typically gauged values. In the present study, the discharge time-series at Tonneins, i.e. the forcing data for the T2D model, is provided either by in-situ observations or by CTRIP simulation, both of which are highly prone to uncertainties.

\begin{figure}[!t]
\centering
\includegraphics[width=0.5\textwidth]{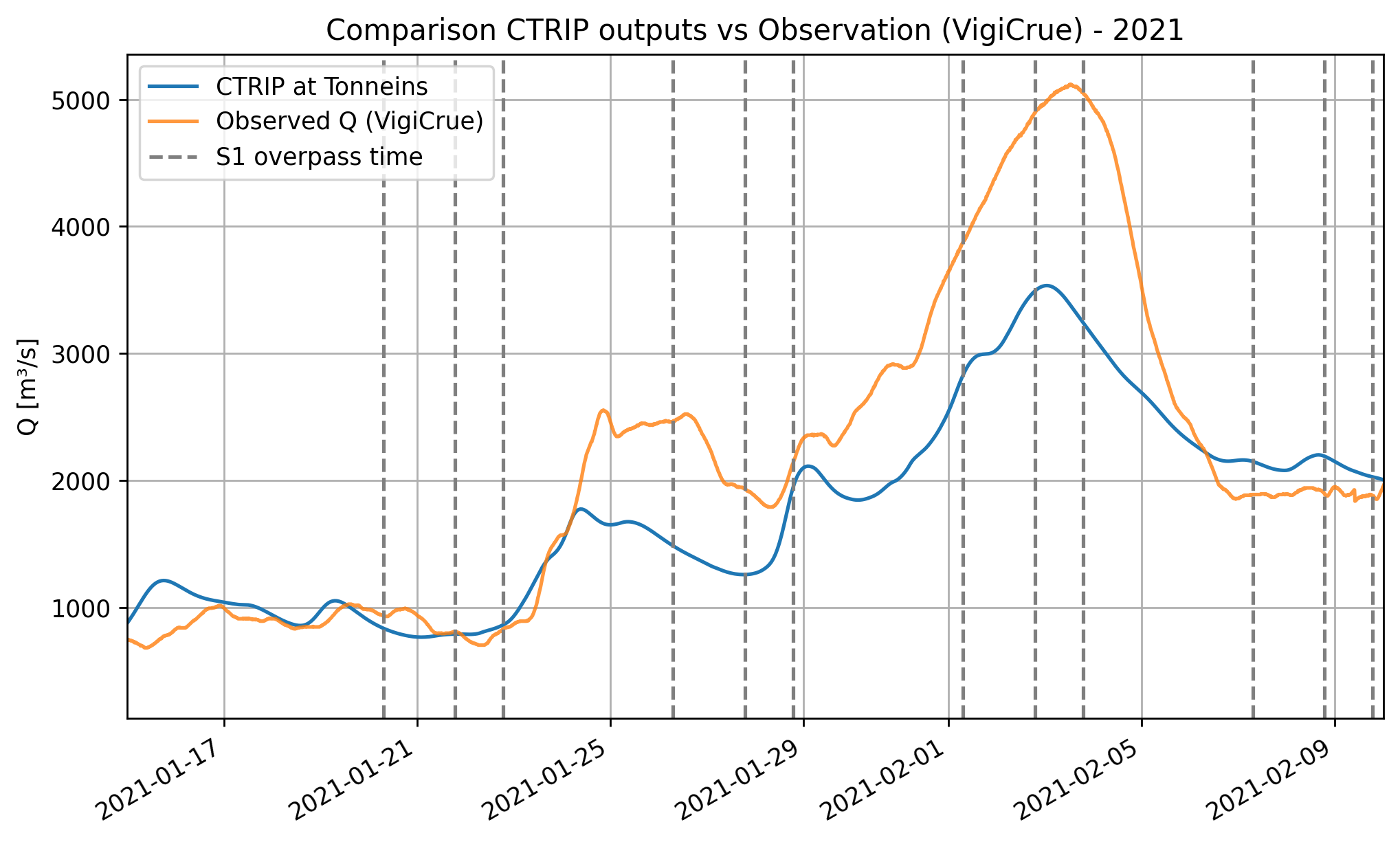}%
\label{sfig:CTRIP_2021}
\caption{Discharge time-series at Tonneins for the 2021 flood events from VigiCrue gauge station (orange line) and simulated by ISBA-CTRIP (blue line). The overpass times of Sentinel-1 are indicated with vertical dashed lines.}
\label{fig_4}
\end{figure}

It is worth noting that the ISBA-CTRIP hydrologic model generally gives better performance for large basins, and that moderate performance can be expected for the medium-sized Garonne basin at Marmande, especially for high flows. In fact, for large flood events, such a model tends to underestimate the discharge. Figure~\ref{fig_4} shows the CTRIP simulated runoff at Tonneins (blue) along with the observed runoff (orange) for the recent major flood event that occurred in 2021 (the acquisition time of the Sentinel-1 images is indicated by vertical dashed lines). These time-series show that CTRIP underestimates the flood peaks by approximately 30\% of the observed peak discharge (CTRIP predicted around 3,500 \Qunit~with respect to an observed discharge of 5,100 \Qunit).



 \section{Data Assimilation}
 \label{DA}

\subsection{Cycled DA analysis and forecast}
\label{subsect:cycle}

The research work is a continuation of \cite{NguyenTGRS2022,nguyenagu2022}. The DA strategy is similar to that presented in \cite{nguyen2023}, here extended to account for the forcing data provided by the large-scale ISBA-CTRIP hydrologic model in a forecasting configuration. It is assumed that the main sources of uncertainty in the chained hydrologic-hydraulic model lie in the CTRIP simulated runoff, in the T2D friction parameters, as well as in the simulated hydraulic state in the floodplain. The errors in the CTRIP forcing data are accounted for by a multiplicative factor $\mu$ applied to the inflow discharge time-series. This correction factor is constant over a DA cycle and varies between DA cycles. In addition, the errors in the hydraulic state due to the absence of evapotranspiration and soil infiltration in the T2D hydraulic model, are accounted for as a correction to the water level $\delta H_k$ (where $k \in [1, 5]$) over five selected subdomains of the floodplain, represented by the hashed regions in Figure~\ref{fig_3}. Within each of these floodplain subdomains, the hydraulic state correction is applied uniformly, i.e., the same value for each mesh node that belong to the same subdomain. Thus, the DA control space contains seven friction coefficients, a multiplicative forcing factor, and five uniform hydraulic state corrections in the floodplain. The EnKF involves 75 members with perturbed control elements following Gaussian distributions. The ensemble DA assimilates in-situ water level measurements and wet surface ratio (WSR) computed over the floodplain subdomains \cite{nguyenagu2022}. This ratio comes directly from observed flood extent maps, derived from SAR images processed by a random forest algorithm from Sentinel-1 images. It is the ratio of the number of wet pixels in the floodplain subdomain to the total number of pixels in the subdomain. 

The EnKF algorithm is favored in this work as it allows the stochastic estimation of the covariance matrices between the model inputs/parameters and outputs without formulating the tangent linear of the hydrodynamic model. In addition, the uncertainty reduction due to a joint state-parameter sequential correction by an EnKF is also enhanced by a Gaussian anamorphosis (GA) \cite{simon2009,simon2012}. This GA algorithm allows us to deal with non-Gaussian errors associated with WSR observations by mapping these non-Gaussian measurements onto a Gaussian space \cite{nguyen2023}. This favors the preservation of the optimality of the performed EnKF. The water level time-series at the three VigiCrue observing stations (i.e. Tonneins, Marmande and La Réole) and the WSR computed over the five floodplain subdomains are assimilated with the EnKF algorithm implemented on the T2D Garonne Marmandaise model. This allows a sequential correction of the friction, the inflow discharge and the water level in the floodplain subdomains.

\begin{figure}[!t]
 \centering
    \includegraphics[width=0.5\textwidth]{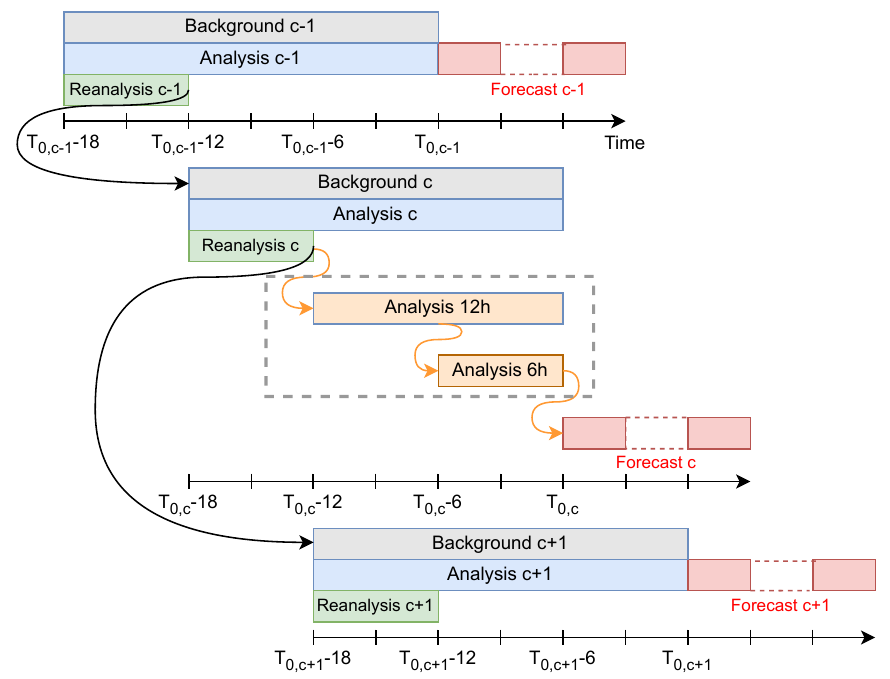}
    \caption{DA workflow for reanalysis (green blocks) and forecast (red blocks) at cycle $c-1$, $c$, and $c+1$.  \fct}
    \label{fig:DA_fct}
\end{figure}

The cycled DA workflow is illustrated in Figure~\ref{fig:DA_fct}. It is implemented for a real-time forecasting system with DA cycles of 18~hr with 6~hr of overlap. Each cycle $c$ consists of a reanalysis phase (green blocks) and a forecast phase (red blocks).
The forecast phase may or may not be activated for a cycle; it is has no impact on the cycled analysis.
The analysis at cycle $c$ assimilates available observations over an 18-hr window up to the present time $T_{0,c}$. This window length was selected to fully take advantage of the RS data that are sparse in time. The resulting corrected friction, forcing and corrective terms to the hydraulic state obtained from the 18-hr analysis window at cycle $c$ are used to run a 6-hr updated simulation (green blocks, between $[T_{0,c}-18; T_{0,c}-12]$) that represents the hindcast reanalysis of the event. The next DA cycle $c+1$ starts from the end of the 6-hr reanalysis at $T_{0,c+1}-18$ (i.e. $T_{0,c}-12$).

The forecast framework is embedded in the cycled DA procedure. When activated for cycle $c$ (e.g. during a flood rising limb), it articulates as three nested assimilations over $c$, first of which is the previously mentioned (blue block). 
Taking into account only observations up to the present $T_{0,c}$, and in order to benefit from sparse RS observations while being consistent with the most recent observations, the 18-hr analysis is followed by a 12-hr and a 6-hr analyses.
These two subsequent assimilations stand in a background step (not shown in Figure~\ref{fig:DA_fct}) and an analysis (orange blocks). These forecast-dedicated steps are framed in the gray-dashed rectangle (shown for cycle $c$ only). They act as external loops (gradually taking into account non-linearity in the filter) and provide the optimal initial state and parameters for the forecast simulation starting at $T_{0,c}$ and lasting for 36~hr. 
Three strategies for forcing are investigated here: (i) using CTRIP predicted runoff for analysis and forecast, called CTRIP-CTRIP (or CC) in the following, (ii) using observed discharge for analysis and CTRIP for forecast, called VigiCrue-CTRIP (or VC) in the following, and (iii) using observed discharge for analysis with a persistent value for forecast, called VigiCrue-constant Q (or VQ) in the following. These strategies for the forecast experiments are gathered in Table~\ref{tab:expe}. It should be noted that in the following, the ensemble and its mean state are used to evaluate the performance of the DA method in forecast mode.

\begin{table}[!t]
\caption{Acronym of the experiments with different forcing strategies in forecast.}
\label{tab:expe}
\centering
\begin{tabular}{|c||c||c|}
\hline
\diagbox{Reanalysis}{Forecast}& CTRIP& Constant Q  \\\hline
CTRIP & CC& - \\\hline
VigiCrue  & VC & VQ \\\hline
\end{tabular}
\end{table}

\subsection{Description of the EnKF algorithm}\label{subsect:EnKF}



In the following, $i \in [1, N_e]$ denotes the member index within an ensemble of size $N_e$. The vector ${\bf x}^{b,i}_c$ (or ${\bf x}^{a,i}_c$) represents the background (or analysis) control vector for member $i$ over a DA cycle $c$. The EnKF background step (gray blocks in Figure~\ref{fig:DA_fct}) consists in propagating the $N_e$ control and model state vectors (i.e. ${\bf x}^{b,i}_{c}$ and ${\bf s}^{b,i}_{c}$, respectively) in time over $[T_{0,c}-18; T_{0,c}]$ with $T_{0,c}$ being the present time for cycle $c$. Then, the EnKF analysis step updates the control ${\bf x}_c^{a,i}$ and computes the associated hydraulic state vector ${\bf s}_c^{a,i}$ over $[T_{0,c}-18; T_{0,c}]$ (blue blocks in Figure~\ref{fig:DA_fct}). The first 6~hr of this integration (green blocks over $[T_{0,c}-18; T_{0,c}-12]$) is the contribution from  cycle $c$ to the reanalysis for the simulated event in hindcast mode.

In the EnKF background step, the background hydraulic state for each member ${\mathbf{s}}^{b,i}_{c}$ results from the integration of the hydrodynamic model ${\cal{M}}_c$: ${\mathbb{R}}^n \rightarrow {\mathbb{R}}^m$  from the control space to the model state (of dimension $m$) over cycle $c$: 
\begin{equation}
{\mathbf{s}}^{b,i}_{c} = {\cal{M}}_{c}\left({\bf s}^{a,i}_{c-1},{\bf x}^{b,i}_{c}\right),
\label{eq:stateforecast}
\end{equation}
where ${\bf s}^{a,i}_{c-1}$ is a restart file saved from the previous analysis at cycle $c-1$, and ${\bf x}^{b,i}_{c}$ is the background control vector at the current cycle. Note that a short spin-up integration is run over the 3~hr preceding each 6-hr window, in order to avoid inconsistencies between the states and the analyzed set of parameters.

The observation vector is denoted by $\mathbf{y}^{o}_c$, it gathers in-situ WL and S1-derived WSR observations for cycle $c$. 
To perform a stochastic EnKF \cite{asch2016data}, the observation vector $\mathbf{y}^{o}_c$ is perturbed by adding $\boldsymbol{\epsilon}_c$ to generate an ensemble of observations $\mathbf{y}^{o,i}_c $.
The observation error $\boldsymbol{\epsilon}_c \sim {\cal N}({\bf 0},{\bf R}_c)$, where ${\bf R}_c= {\sigma^2_{obs}} {\mathbf I}_{n_{obs}}$ is the observation error covariance matrix, assumed to be diagonal of standard deviation $\sigma_{obs}$ as the observation errors are assumed to be uncorrelated Gaussian.
It should be noted that within a cycle, the observation errors decreases as the observation time gets closer to the present time so that the recent observations weight more in the analysis than the older ones (i.e. recent observations are more reliable). 
The equivalent of the model in the observation space, denoted by ${\bf y}^{b,i}_c$, is computed with the observation operator ${{\cal{H}}_c}$: ${\mathbb{R}}^m \rightarrow {\mathbb{R}}^{n_{obs}}$: 
\begin{equation}
{\mathbf{y}}^{b,i}_{c} = {{\cal{H}}_c}\left({\mathbf{s}}^{b,i}_{c}\right).
\label{eq:ctlequivobs}
\end{equation}

Next, the EnKF analysis step updates the control ${\bf x}_c^{a,i}$ and the associated model state vector ${\bf s}_c^{a,i}$ in an anamorphosed space \cite{nguyen2023} using the transformed observation operator ${{\tilde{\cal{H}}}_c}$:
\begin{equation}
{\bf x}_c^{a,i} = {\bf x}^{b,i}_{c} + \mathbf{K}_{c} \left(\tilde{\bf y}^{o,i}_{c} - \tilde{\bf y}^{b,i}_{c}\right).
\label{eq:ctlana}
\end{equation}
The Kalman gain is further computed from covariance matrices that are stochastically estimated within the ensemble, considering anamorphosed observation vectors  $\tilde{\bf y}^{b,i}$:
\begin{equation}
	\mathbf{K}_c = \mathbf{P}^{\bf{x},\tilde{\bf{y}}}_c {\left[ \mathbf{P}^{\tilde{\bf{y}},\tilde{\bf{y}}}_c + \mathbf{R}_{c} \right]}^{-1}.
	\label{eq:EnKF_ana_Klambda_gain_chap12}
\end{equation}

Similar to Eq.~\eqref{eq:stateforecast},  the hydrodynamic state ${\bf s}^{a,i}_c$ associated with each analyzed control vector ${\bf x}^{a,i}_c$  results from the integration of the hydrodynamic model ${\cal{M}}_c$ with the updated parameters in ${\bf x}^{a,i}_{c}$: 
\begin{equation}
	{\mathbf{s}}^{a,i}_{c} = {\cal{M}}_{c}\left({\bf s}^{a,i}_{c-1},{\bf x}^{a,i}_{c}\right),
	\label{eq:stateanalyzed}
\end{equation}
where ${\bf x}^{a,i}_{c}$ gathers $(K_{s_k})^{a,i}_{c}, \mu^{a,i}_c$ and the state correction in the floodplain subdomains $\delta H^{a,i}_k$ over $W_c$, starting from the same initial condition as each background simulation within the ensemble. It should be noted that the ensemble mean $\overline{\delta H_k^a}$ is used in place of $\delta H^{a,i}_k$ in order to preserve a smooth WL field.

When activated, the forecast step unfolds as follows.
For each member, the first (respectively, second) forecast-dedicated analysis, in gray-dashed rectangle, uses the previous 18-hr (respectively, 12-hr) analyzed control vector as background from which a new shortened background trajectory is issued over $[T_{0,c}-12; T_{0,c}]$ (respectively, over $[T_{0,c}-6; T_{0,c}]$). Note that these shortened background simulations are not represented in Figure~\ref{fig:DA_fct} for clarity purposes. Each of them is followed by an analysis (orange blocks) similar to Eq.~\eqref{eq:ctlana}. The control vector resulting from the last 6-hr analysis, denoted ${\bf x}^{a,i}_{c,fT_0}$, are then used to integrate an ensemble of forecast simulations starting from the ensemble of hydraulic states at $T_{0,c}$, noted ${\bf s}^{a,i}_{c, fT_0}$, over 36~hr (red blocks in Figure~\ref{fig:DA_fct}):
\begin{equation}
	{\mathbf{s}}^{f,i}_{c} = {\cal{M}}_{c}\left({\bf s}^{a,i}_{c, fT_0},{\bf x}^{a,i}_{c,fT_0}\right).
	\label{eq:stateanalyzed}
\end{equation}
\sophie{The ensemble of forecast trajectories ${\mathbf{s}}^{f,i}_{c}$ for $i \in [1, N_e]$ is further compared to the reanalysis (green blocks) and observations in order to assess the predictive capabilities of the DA workflow in real-time.}

\section{Results}
\label{results}
Three types of experiment are carried out for the 2021 flood event over the studied Garonne Marmandaise catchment: a free run  without assimilation, also known as open-loop (OL), a DA experiment (named IDA) that assimilates only in-situ water-level, and a DA experiment (named IGDA) that assimilates both in-situ water-level and remote-sensing WSR observations. It should be noted that the 2021 flood event was not used for model calibration. In order to issue the reanalysis and assess them for this flood event in a hindcast mode, each simulation is carried out using either forcing at Tonneins from VigiCrue (experiments denoted by $\mathrm{OL}^\mathrm{V}$, $\mathrm{IDA}^\mathrm{V}$ and $\mathrm{IGDA}^\mathrm{V}$), or CTRIP forcing (experiments denoted $\mathrm{OL}^\mathrm{C}$, $\mathrm{IDA}^\mathrm{C}$, and $\mathrm{IGDA}^\mathrm{C}$).  
\sophie{Then. the forecast mode is carried out from $\mathrm{IGDA}$ re-analysis as shown in Figure~\ref{fig:DA_fct} using one of the three forcing strategies presented in Table~\ref{tab:expe}.}

Reanalysis results are first assessed in the control space. The \sophie{reanalysis and forecast results are then assessed} in the observation space using 1D and 2D metrics. In 1D assessment, the RMSE between observed and simulated water level at in-situ observing stations is computed to assess the dynamics of the flow in the riverbed. On the other hand, in the 2D assessment, the agreements between flood extents simulated with T2D and flood extents derived from Sentinel-1 images are shown with a contingency map and an overall Critical Success Index (CSI) to assess the dynamics of the flow in the floodplains. The reanalysis and forecast results are finally compared to independent the High Water Marks (HWM) dataset.


\subsection{Reanalysis results in the control space}
\begin{figure*}[!t]
\centering
\includegraphics[width=0.9\textwidth]{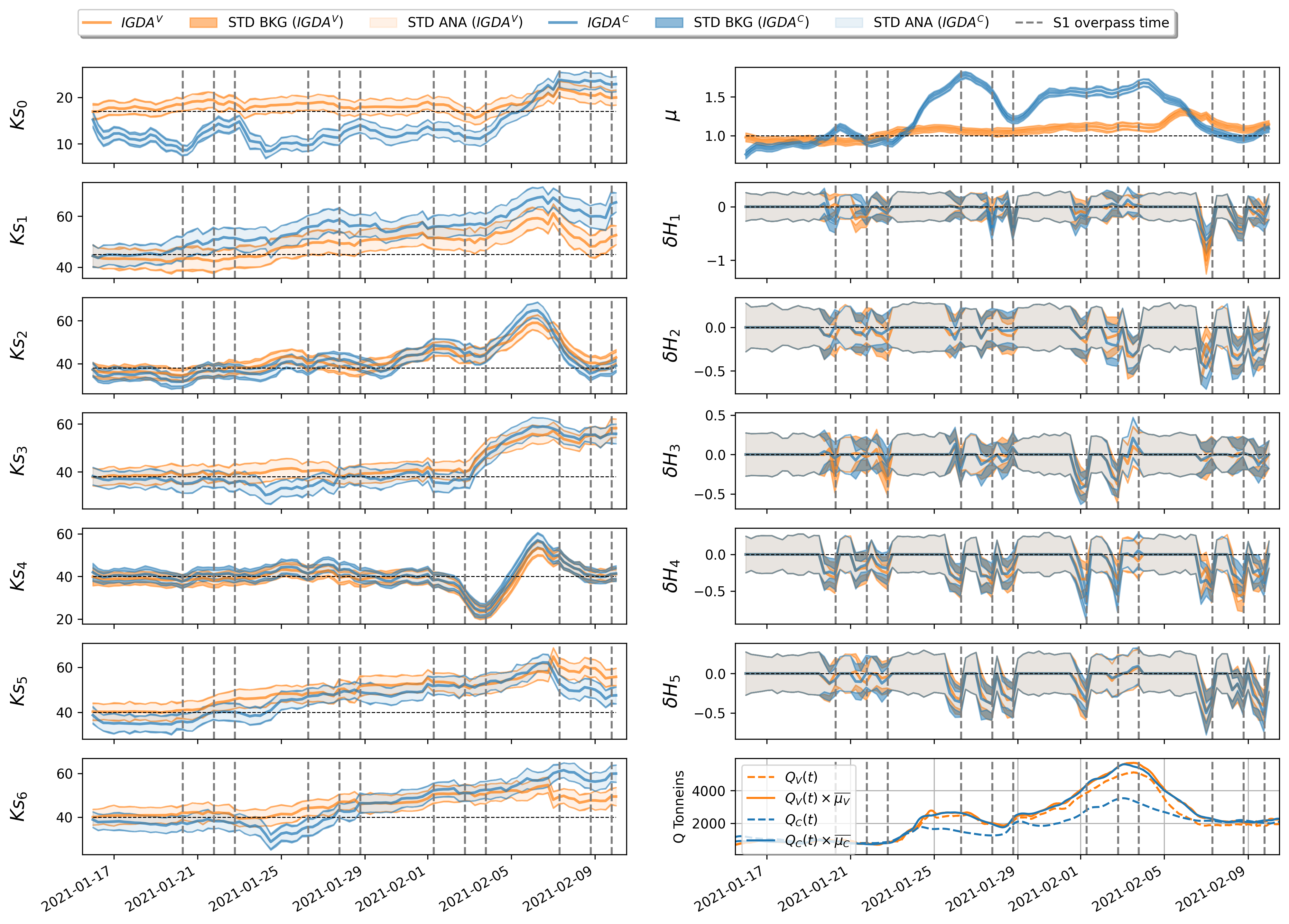}
\caption{DA analysis (IGDA) for 2021 flood event, in the control space for friction coefficient $K_{s_0}$ in the floodplain (left column, top panel), $K_{s_k}$ with $k \in [1, 6]$ in the riverbed (left column), multiplicative coefficient $\mu$ for inflow discharge at Tonneins (right column, top panel), uniform water level correction $\delta H_k$ (with $k \in [1, 5]$, right column) and reconstructed inflow discharges at Tonneins (right column, bottom panel).} 
\label{fig_5}
\end{figure*}

Figure~\ref{fig_5} shows the analyzed parameters from the different DA reanalyses forced by VigiCrue or CTRIP, with solid orange and blue lines respectively representing the mean of the control vector analysis for $\mathrm{IGDA}^\mathrm{V}$ and $\mathrm{IGDA}^\mathrm{C}$, for the 2021 flood event. 
The shaded envelopes represent the standard deviation of the ensemble (solid color for background and light color for analysis). These envelopes overlap on Figure~\ref{fig_5}, and appear in a gray color when all four envelopes overlap. The default values are indicated by horizontal dashed lines.  The analyzed values for the friction coefficients $K_{s_k}$ (with $k \in [0, 6]$) are shown on the left column. The inflow multiplicative correction $\mu$ is shown on the right column along with the analyzed values of the hydraulic state correction $\delta H_k$ (with $k \in [1, 5]$). The bottom right panel displays the reconstructed upstream discharges using the mean of the $\mu$ analysis for the ensemble. The overpass times of Sentinel-1 over the 2021 flood event are indicated by vertical dashed lines.

First and foremost, the DA analysis for $\mathrm{IGDA}^\mathrm{V}$ and $\mathrm{IGDA}^\mathrm{C}$ provides similar corrections to the friction coefficients of the riverbed (except $K_{s_1}$). Yet, the analyses for the friction in the floodplain $K_{s_0}$ and the correction to the inflow at Tonneins are very different. Indeed, CTRIP-simulated runoff is significantly under-predicted compared to the observed discharge and leads to low water levels in the whole catchment without DA (as shown below in Figure~\ref{fig_7}). As a consequence, the EnKF prescribes a multiplicative factor $\mu > 1$ to account for this lack of runoff, especially during the high flow between 23/01/2021 and 06/02/2021. 
During the flood recess after 07/02/2021, CTRIP slightly exceeds the VigiCrue observations. Yet, for some equifinality reasons, the multiplicative factor $\mu$ in $\mathrm{IGDA}^\mathrm{C}$ remains close to 1  and the overestimation is accounted for by correction in $K_{s_0}$, $K_{s_1}$, $K_{s_5}$, and $K_{s_6}$ as well as by the  hydraulic state correction $\delta H_k$ (with $k \in [1, 5]$) in the floodplain subdomains.

Since the velocity in the floodplain is small, the friction coefficient in the floodplain ($K_{s_0}$) only has a minor impact on the  dynamics of the flow. This leads to small correction of $K_{s_0}$ for $\mathrm{IGDA}^\mathrm{V}$. Yet, when the misfit to the observations is large (i.e. when T2D is forced with the underestimated CTRIP runoff in $\mathrm{IGDA}^\mathrm{C}$), the correction to $K_{s_0}$ is large.  $K_{s_0}$ is reduced below the default value, thus increasing the  water levels. 
Reversely, the friction in the flood plain is reduced with higher $K_{s_0}$ during recess, to decrease the water level in the floodplain that T2D struggles to empty after the flood peak  due to the absence of ground infiltration and evapotranspiration processes in the model. 

The hydraulic state corrections $\delta H_k$ (with $k \in [1, 5]$) are shown with non-null values at each S1 overpass. The corrections are mostly negative acting for the evacuation of water after the flood peaks. 
The reconstructed hydrograph at Tonneins for $\mathrm{IGDA}^\mathrm{C}$,  is close to that of $\mathrm{IGDA}^\mathrm{V}$ which illustrates that the EnKF succeeds in retrieving a realistic forcing time-series even when the a priori runoff from CTRIP is significantly underestimated. 


\subsection{Reanalysis results in the observation space} \subsubsection{1D metrics on water levels at observing stations}
 
\begin{figure*}[!t]
\centering
\subfloat[$\mathrm{OL}^\mathrm{V}$]{\includegraphics[width=0.75\textwidth]{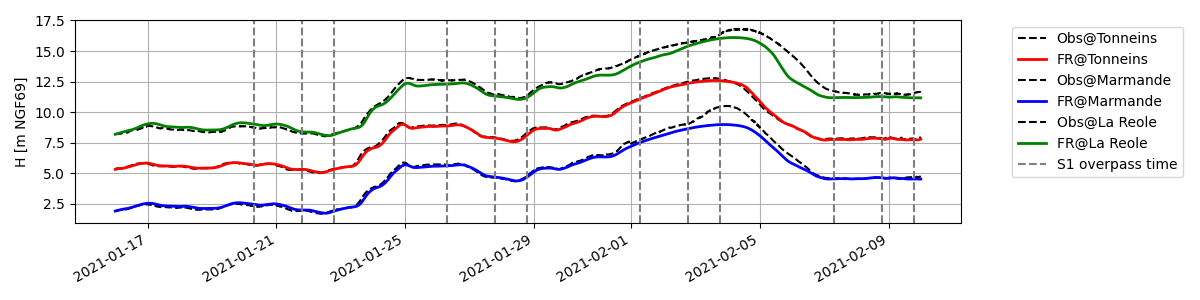}%
\label{fig_6a}}

\subfloat[$\mathrm{IGDA}^\mathrm{V}$]{\includegraphics[width=0.75\textwidth]{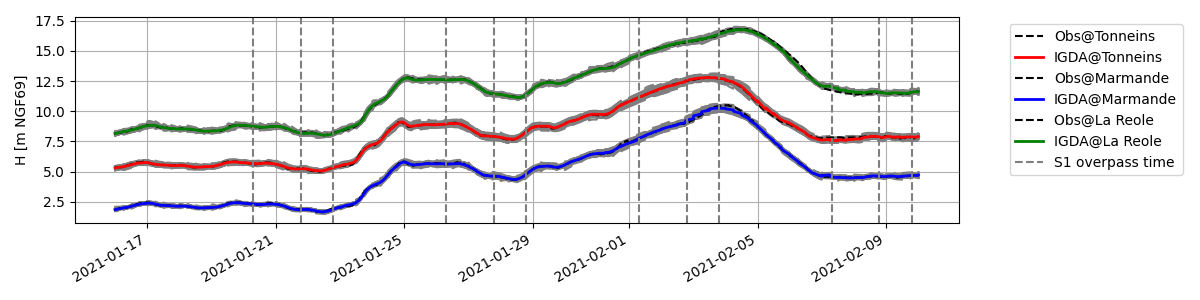}%
\label{fig_6b}}
\caption{Water levels at the observing stations at Tonneins (red), Marmande (blue) and La Réole (green) simulated by (a) $\mathrm{OL}^\mathrm{V}$ and  (b) $\mathrm{IGDA}^\mathrm{V}$, compared to the observed water levels (plotted in black-dashed lines), for the experiments forced by VigiCrue discharge. \reana}
\label{fig_6}
\end{figure*}

Figure~\ref{fig_6} shows the water levels at the observing stations at Tonneins (red), Marmande (blue) and La Réole (green) simulated by $\mathrm{OL}^\mathrm{V}$ (top panel) and $\mathrm{IGDA}^\mathrm{V}$ (bottom panel) compared to the observed water levels (black-dashed lines), for experiments forced by VigiCrue time-series. It should be noted that $\mathrm{OL}^\mathrm{V}$  slightly underestimates the flood peak. 
For $\mathrm{IGDA}^\mathrm{V}$, the ensemble of analysis is plotted in gray and the mean of the analysis is plotted in color. This result shows the efficiency of the DA strategy to correct the flow dynamics in the riverbed when forced by observed runoff.
 
\begin{figure*}[!t]
\centering
\subfloat[$\mathrm{OL}^\mathrm{C}$]{\includegraphics[width=0.75\textwidth]{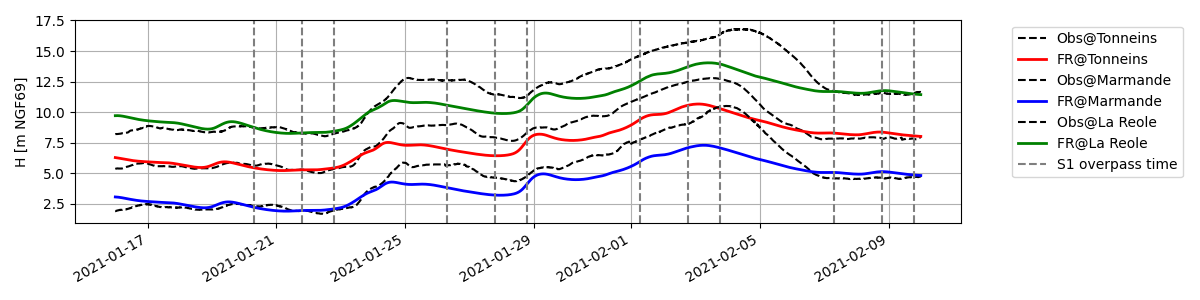}%
\label{fig_7a}}

\subfloat[$\mathrm{IGDA}^\mathrm{C}$]{\includegraphics[width=0.75\textwidth]{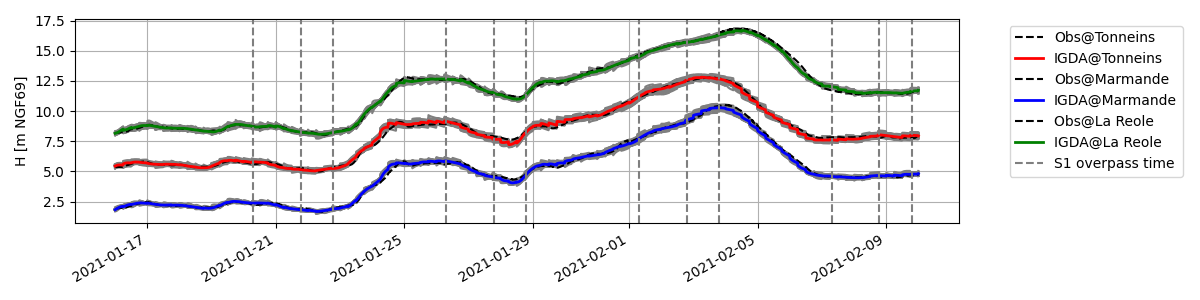}%
\label{fig_7b}}
\caption{Water levels at the observing stations at Tonneins (red), Marmande (blue) and La Réole (green) simulated by (a) $\mathrm{OL}^\mathrm{C}$  and (b) $\mathrm{IGDA}^\mathrm{C}$ (bottom panel), compared to the observed water levels (plotted in black-dashed lines), for the experiments forced by CTRIP runoff. \reana}
\label{fig_7}
\end{figure*}

Figure~\ref{fig_7} is similar to Figure~\ref{fig_6} but with CTRIP runoff. The underestimation of the peak is all the more visible for the simulation forced by CTRIP than by VigiCrue. Underestimated CTRIP runoff time-series leads to a significant underestimation of water level at high-water period, including the flood peak in $\mathrm{OL}^\mathrm{C}$. All four DA experiments lead to significant correction on the control leading to improved water levels at all three observing stations over the entire event. This illustrates how DA improves simulated water levels in the river, even when the input forcing is only a coarse approximation of the real inflow. 
 
Table~\ref{tab:table1} summarizes the 1D quantitative results for all six experiments. The RMSEs between the simulated and the observed water levels at each observing station are computed over time for the entire 2021 flood event. Coherent to the water level plots displayed above, the RMSEs underline the underestimation problem when CTRIP runoff is used without DA, resulting in errors greater than 1 meter, which are critical regarding flood hazard assessments. Moreover, for both VigiCrue and CTRIP runoff, the assimilation of in-situ observations only in IDA as well as the joint assimilation of in-situ water level and RS-derived WSR observations in IGDA leads to a significant improvement with reduced RMSE values. Indeed, the gain with respect to OL amounts to 65.15\% when using observed discharge and 88.69\% when using CTRIP runoff for $\mathrm{IGDA}^\mathrm{C}$, with very similar results for $\mathrm{IDA}^\mathrm{C}$. This demonstrates that the assimilation of in-situ data only, efficiently constrains the assimilation algorithm. The RMSEs at observing stations from the CTRIP-forced experiment $\mathrm{IGDA}^\mathrm{C}$ remain slightly larger (14-17 cm) than those of the simulations forced by VigiCrue discharges (7-9 cm). 

\begin{table}[!t]
\caption{Water level RMSE computing at observing stations with respect to in-situ observed water levels. \reana}
\label{tab:table1}
\centering
\begin{tabular}{|c||c||c||c||c|}
\hline
& \multicolumn{4}{c|}{RMSE [m]}\\\hline
& Tonneins & Marmande & La Réole & Gain \\\hline
$\mathrm{OL}^\mathrm{V}$ & 0.106 & 0.392 & 0.536 & - \\\hline
$\mathrm{IDA}^\mathrm{V}$ & 0.062 & 0.071 & 0.081 & 69.43\% \\\hline
$\mathrm{IGDA}^\mathrm{V}$ & 0.073 & 0.074 & 0.090 & 65.15\% \\\hline
 &  &  &  &  \\\hline
$\mathrm{OL}^\mathrm{C}$ & 1.209 & 1.405 & 1.598 & - \\\hline
$\mathrm{IDA}^\mathrm{C}$ & 0.160 & 0.148 & 0.130 & 89.37\% \\\hline
$\mathrm{IGDA}^\mathrm{C}$ & 0.166 & 0.160 & 0.141 & 88.69\% \\
\hline
\end{tabular}
\end{table}
	
\subsubsection{2D metrics on flood extent over the simulation domain, using with contingency maps and Critical Success Index (CSI)}

\begin{figure*}[!t]
\centering
\includegraphics[width=\textwidth]{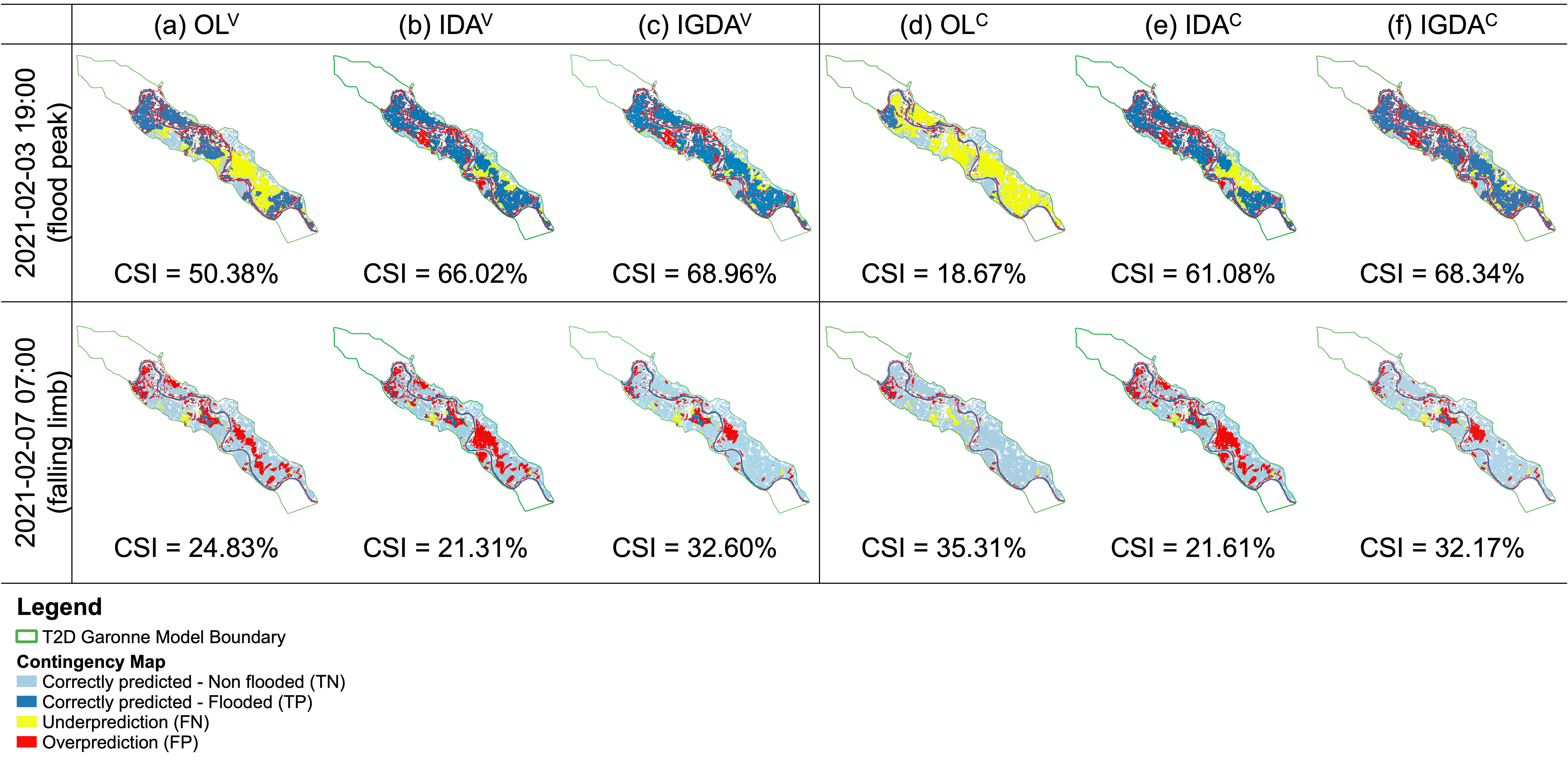}
\caption{Contingency maps and CSI computed between simulated flood extents and S1-derived observed flood extents for $\mathrm{OL}^\mathrm{V}$, $\mathrm{IDA}^\mathrm{V}$, $\mathrm{IGDA}^\mathrm{V}$, $\mathrm{OL}^\mathrm{C}$, $\mathrm{IDA}^\mathrm{C}$ and $\mathrm{IGDA}^\mathrm{C}$ at the flood peak and at recess time for the 2021 flood event. \reana}
\label{fig_8}
\end{figure*}

The merits of assimilating WSR in addition to in-situ data, is significant in the floodplain. The contingency maps between the simulated flood extent and the observed flood extent derived from Sentinel-1 images are computed and displayed in Figure~\ref{fig_8}. It indicates for each pixel within the studied catchment if the simulation succeeds or fails with respect to the observation. If a pixel is correctly predicted as flooded or correctly identified as non-flooded according to the Sentinel-1-derived flood extent maps, the outcome is labeled as True Positive (TP) or True Negative (TN), respectively. Otherwise, it is labeled as False Positive (FP) or False Negative (FN), whether a pixel is not correctly predicted as flooded or not correctly predicted as non-flooded, respectively.  The CSI, also called Threat Score, computes the ratio between the number of TP over the sum pixels of TP, FP, and FN. A CSI close to 100\% indicates a perfect result with respect to the reference (here S1-derived flood extents). 

At the flood peak on 03/02/2021 (top panel of Figure~\ref{fig_8}), both OL experiments show large areas of underpredicted flooding (shown by yellow areas), with $\mathrm{OL}^\mathrm{C}$ being the more severe case due to CTRIP providing understimated runoff. These underprediction areas are significantly reduced by  DA for both $\mathrm{IGDA}^\mathrm{V}$ and $\mathrm{IGDA}^\mathrm{C}$. It should be noted that the assimilation of in-situ only data in $\mathrm{IDA}^\mathrm{V}$ and $\mathrm{IDA}^\mathrm{C}$ brings some improvement with respect to OL but significant underprediction remains, especially when CTRIP runoff is used. This demonstrates the merits of the assimilation of WSR and the associated correction of the hydraulic state in the subdomains of the floodplain with respect to the in-situ only IDA strategy. 
Despite the understimated CTRIP runoff, the CSI score resulting from $\mathrm{IGDA}^\mathrm{C}$  reaches that of $\mathrm{IGDA}^\mathrm{V}$. As such, $\mathrm{IGDA}$ significantly improves the CSI score for both forcing experiments with CSI of 68-69\%.

During the flood recess on 07/02/2021 (bottom panel of Figure~\ref{fig_8}), both OL experiments tend to over-predict flooding (shown by red areas) as the T2D hydrodynamic model struggles to empty the floodplain after the flood peak. The assimilation of in-situ data in IDA tends to even downgrade the results in the floodplain.  Thus, the merits of assimilating S1-derived WSR become noticeable in $\mathrm{IGDA}$ experiments. They significantly reduce the overflooded areas, thanks to the correction of the water level in subdomains of the floodplain $\delta H_k$ with $k \in [1, 5]$. Thanks to such a reduction, the CSI score resulting from $\mathrm{IGDA}^\mathrm{V}$  is improved, compared to $\mathrm{IDA}^\mathrm{V}$ . When using CTRIP runoff, it can be noted that the CSI of $\mathrm{IGDA}^\mathrm{C}$ (32.17\%) during the flood recession is still lower than that of $\mathrm{OL}^\mathrm{C}$ (35.31\%). Indeed, the lack of inflow discharge results in the water extent remaining small at flood recess. Such an assessment (between $\mathrm{OL}^\mathrm{C}$ and $\mathrm{IGDA}^\mathrm{C}$) should be made alongside with the 1D assessment for the same experiments (in which, RMSEs by $\mathrm{IGDA}^\mathrm{C}$ reduced by 73.69\% compared to $\mathrm{OL}^\mathrm{C}$). A similar CSI between $\mathrm{IGDA}^\mathrm{V}$ and $\mathrm{IGDA}^\mathrm{C}$ (respectively 32.60\% and 32.17\%) indicates the limitations of the implemented DA strategy, regardless of the inflow discharge. This also advocates for further improvement of the model physical processes regarding infiltration and evapotranspiration parameters.

\subsection{Forecast results in the observation space}

\begin{figure*}[!t]
 \centering
    \begin{minipage}{0.32\textwidth}
    
    \subfloat[IGDA using CTRIP-CTRIP]{
    \includegraphics[width=\textwidth]{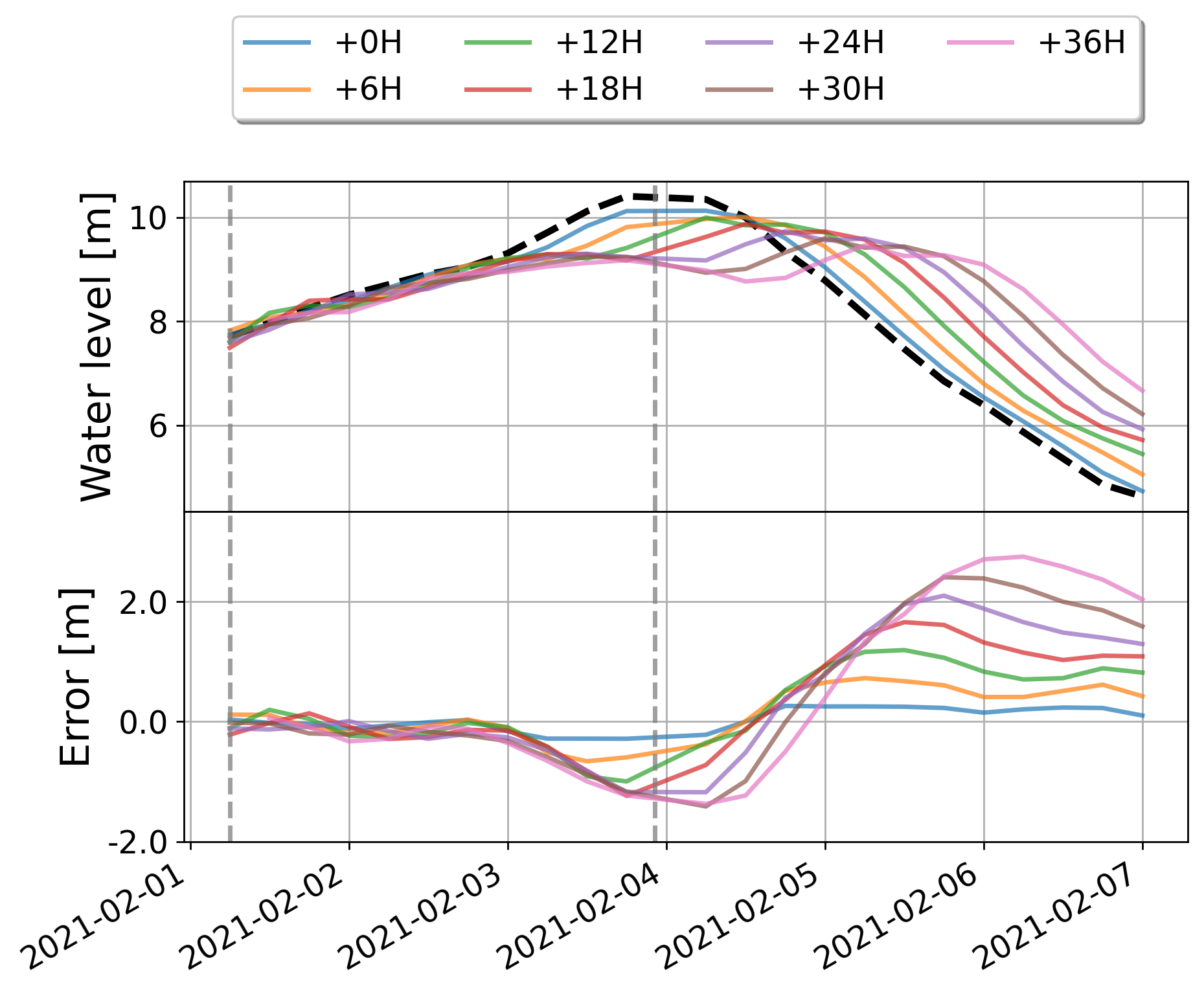}
    }
    
    \end{minipage}
    \hfill
    \begin{minipage}{0.32\textwidth}
    
    \subfloat[IGDA using VigiCrue-CTRIP]{
    \includegraphics[width=\textwidth]{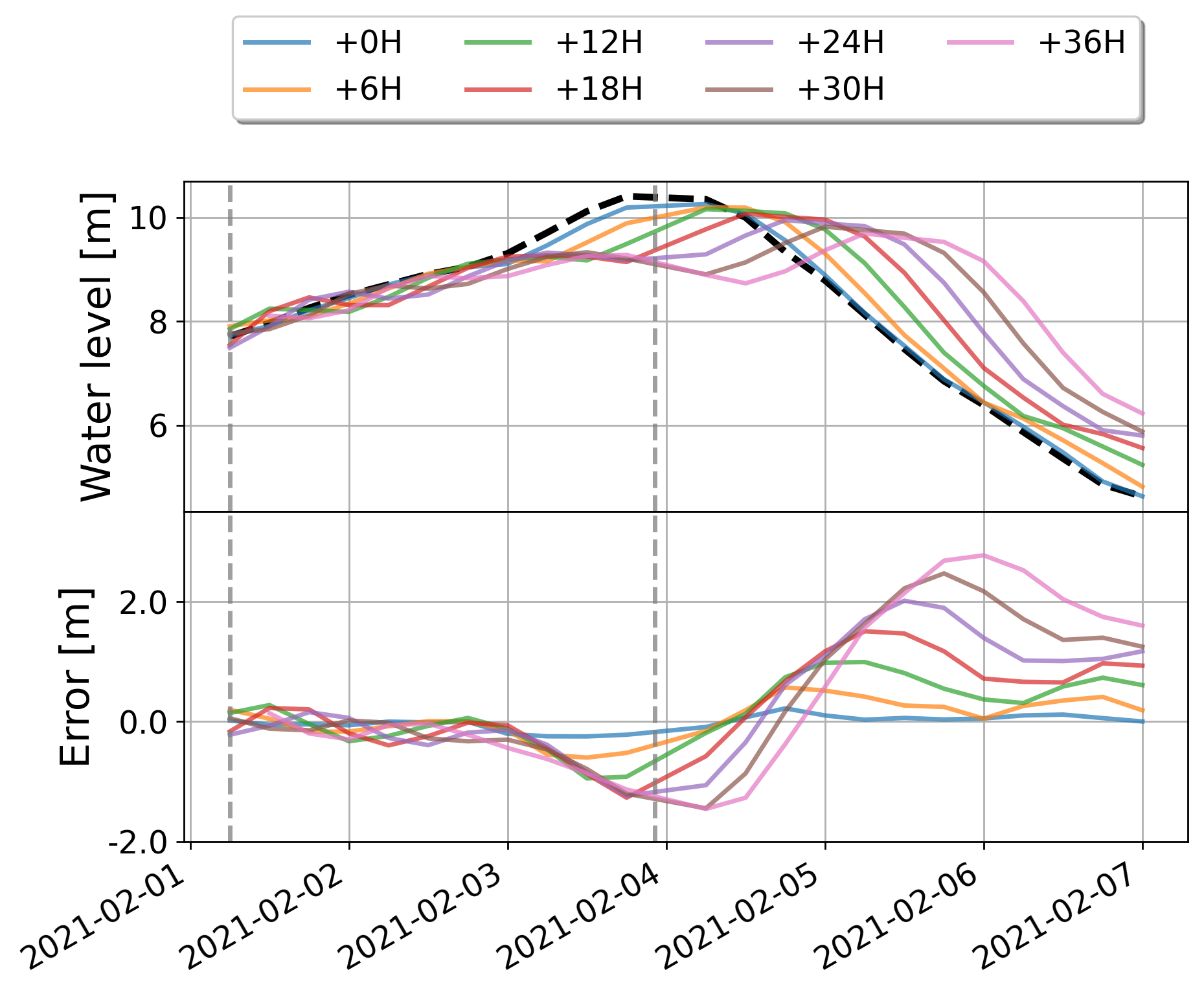}
    }
    
    \end{minipage}
    \hfill
    \begin{minipage}{0.32\textwidth}
    
    \subfloat[IGDA using 
 VigiCrue-constant Q]{
    \includegraphics[width=\textwidth]{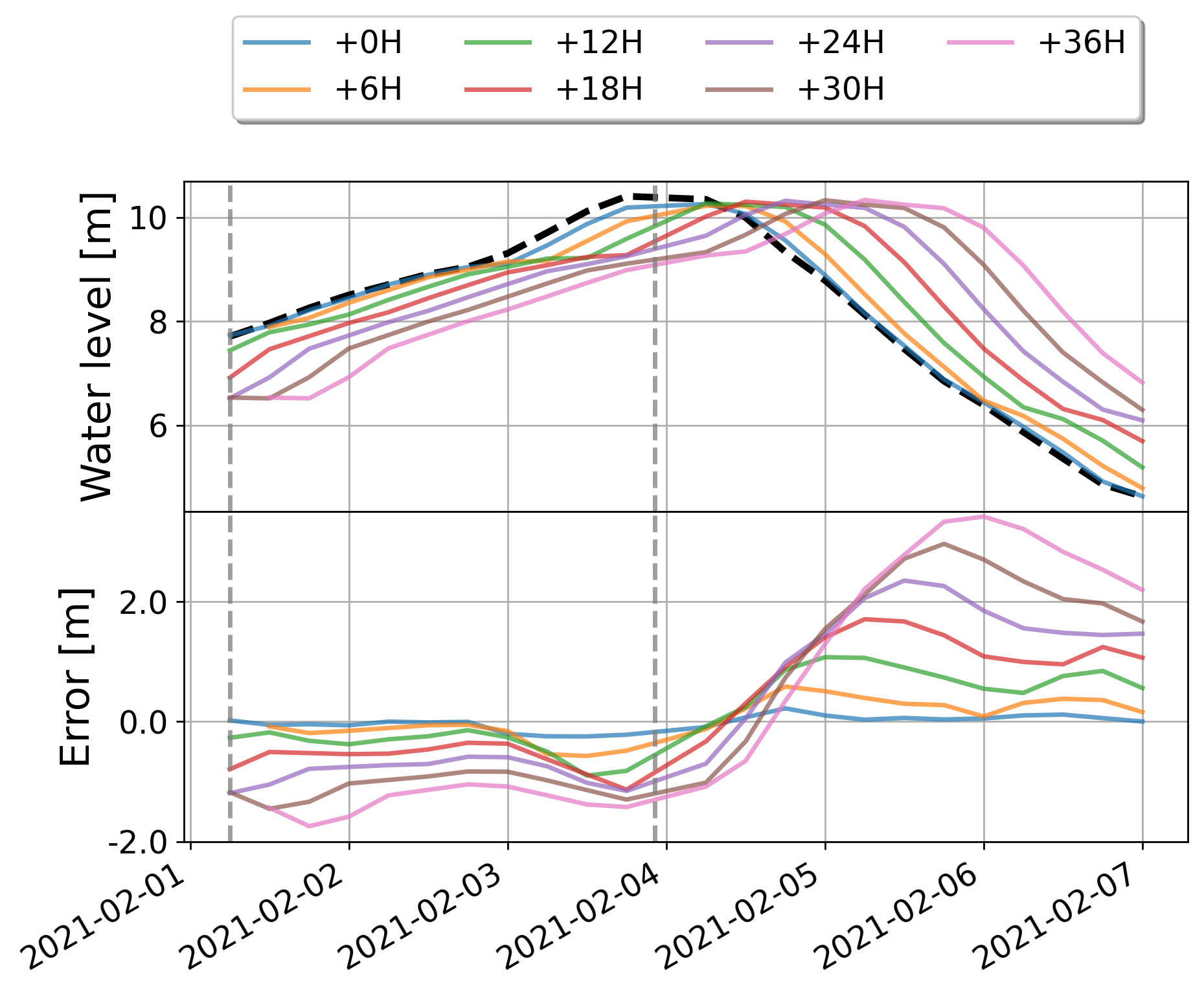}
    }
    
    \end{minipage}
     \caption{Forecasted water levels (averaged among $N_e$ members) at increasing lead times (+0~hr, +6~hr, +12~hr, +18~hr, +24~hr, +30~hr, and +36~hr) in solid colored lines and their errors with respect to the observed water levels in black-dashed line at Marmande observing station. Left column: IGDA using CTRIP-CTRIP; middle column: IGDA using VigiCrue-CTRIP; right column: IGDA using VigiCrue-constant Q. \fct}
    \label{fig:DA2_fct_leadtime}
\end{figure*}

\begin{figure*}[!t]
 \centering
     \subfloat[IGDA using CTRIP-CTRIP]{
    \includegraphics[width=0.32\textwidth]{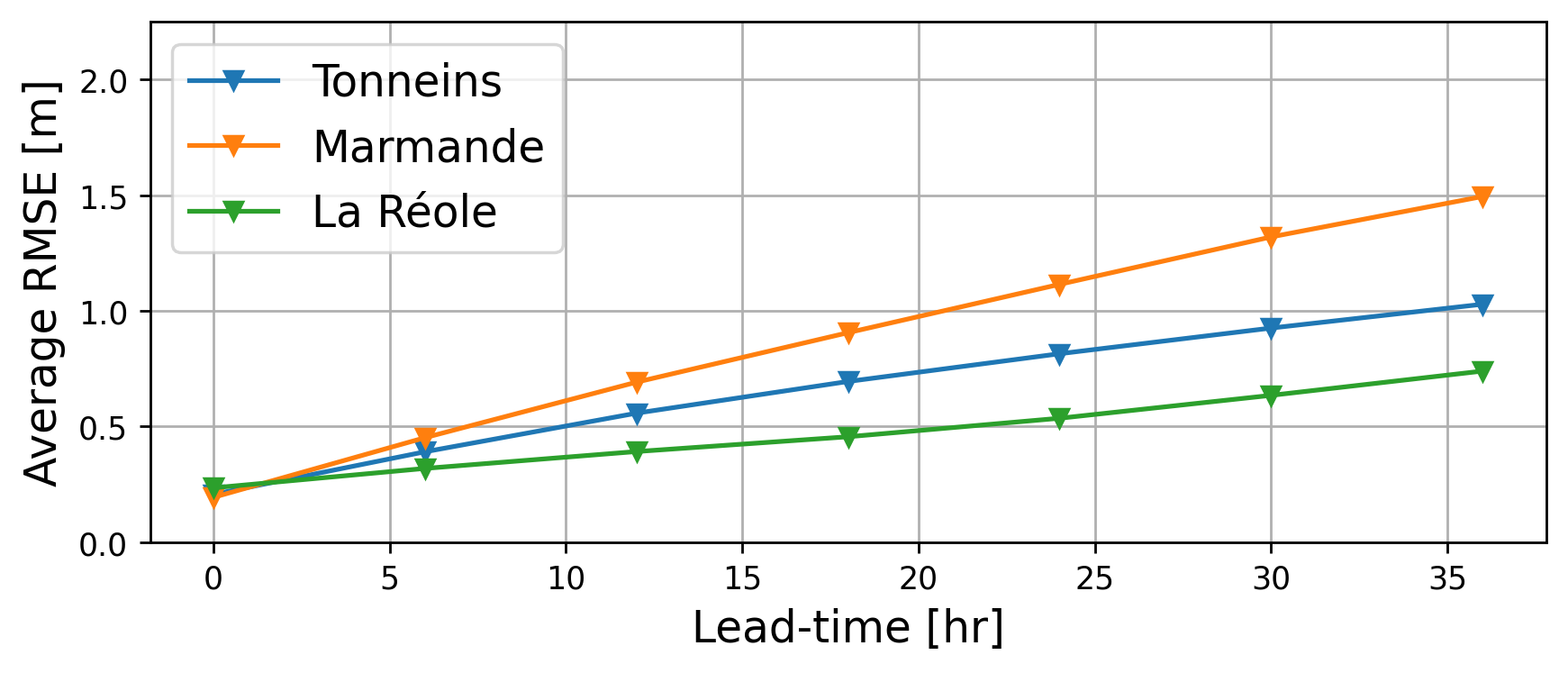}
     }
     \subfloat[IGDA using VigiCrue-CTRIP]{
     \includegraphics[width=0.32\textwidth]{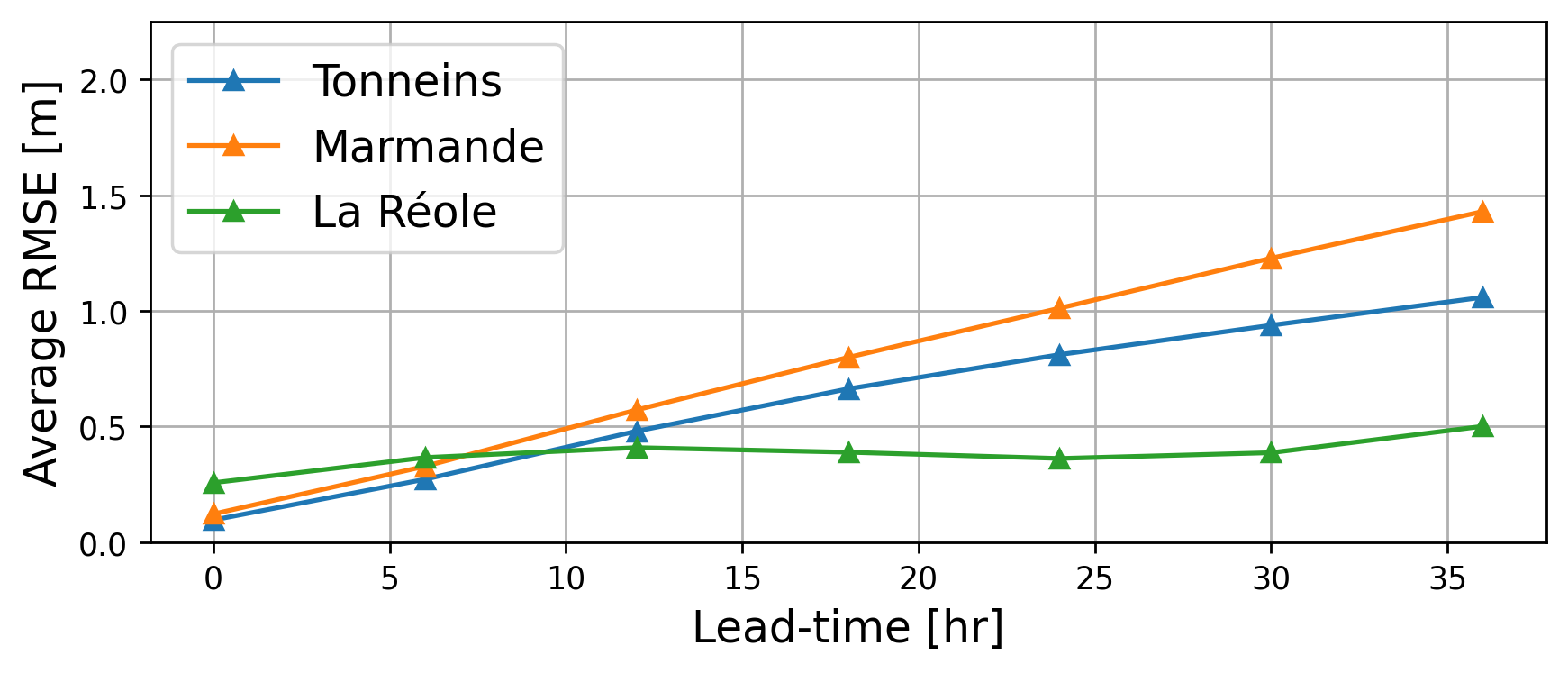}
     }
     \subfloat[IGDA using 
 VigiCrue-constant Q]{
     \includegraphics[width=0.32\textwidth]{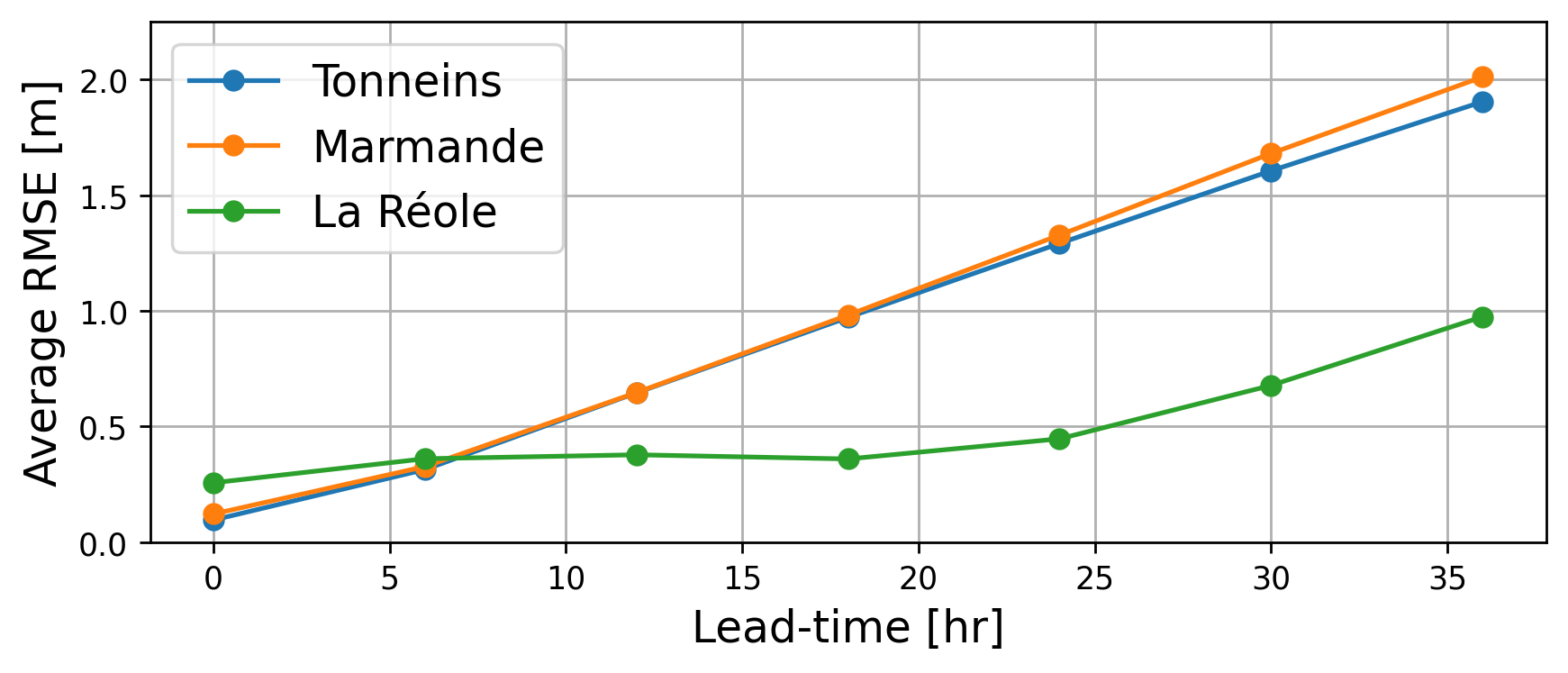}
     }
    \caption{$\mathrm{RMSE}$ between forecasted water level (averaged among $N_e$ members) and observation, computed  over the flood peak period 01/02/2021 to 07/02/2021 at Tonneins (blue line), Marmande (orange line) and La Réole (green line), along increasing lead time. (a) IGDA using CTRIP-CTRIP; (b) IGDA using VigiCrue-CTRIP; (c) IGDA using VigiCrue-constant Q. \fct}
    \label{fig:RMSE_fct}
\end{figure*}

\begin{figure}[!t]
 \centering
    \includegraphics[width=0.45\textwidth]{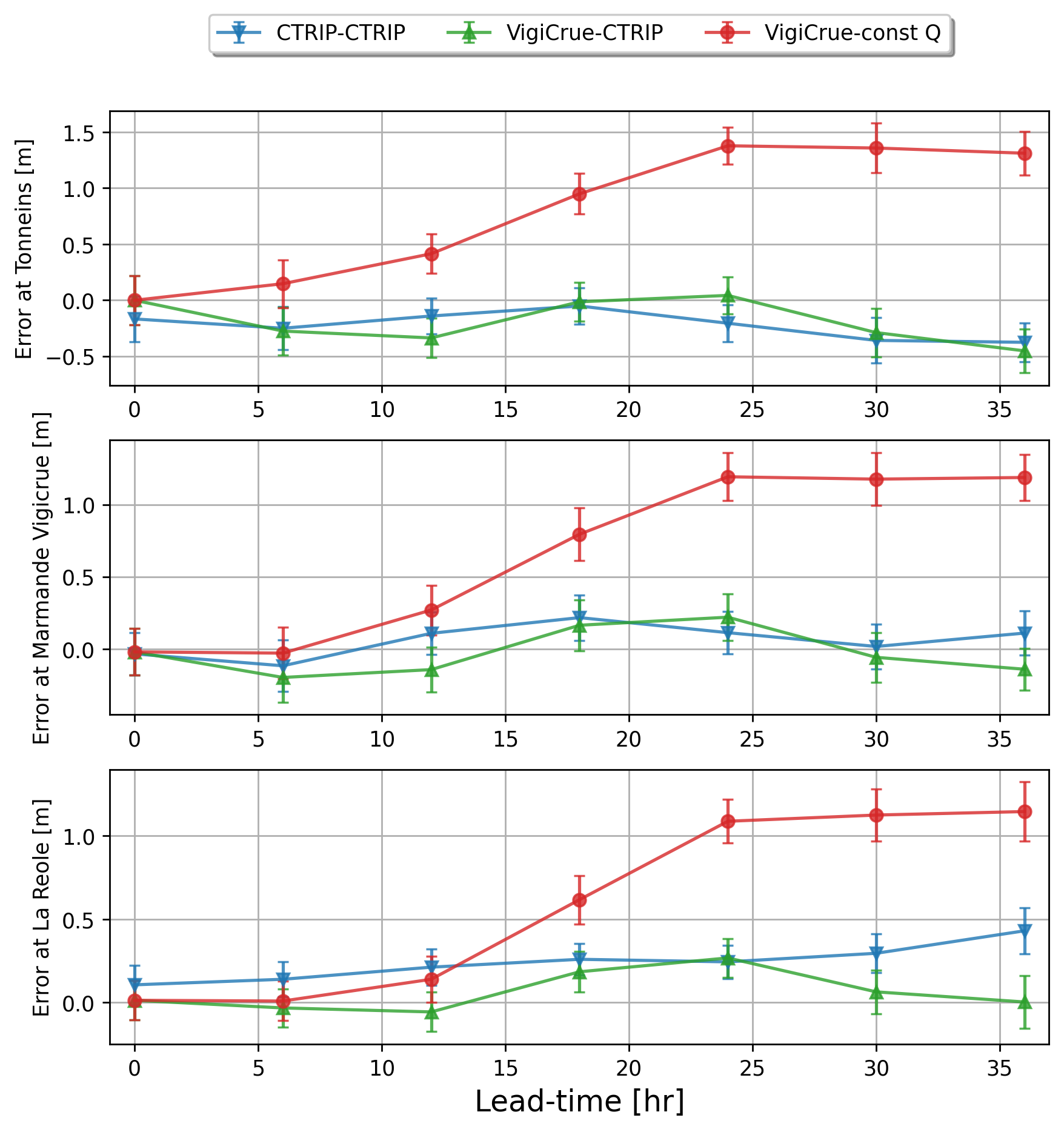}
    \caption{Forecast error at Tonneins, Marmande, and La Réole for the target date 01/02/2021 06:00 (rising limb) with various lead-times. \fct}
    \label{fig:fct_error_targetdate_20210201}
\end{figure}

\begin{figure}[!t]
 \centering
    \includegraphics[width=0.45\textwidth]{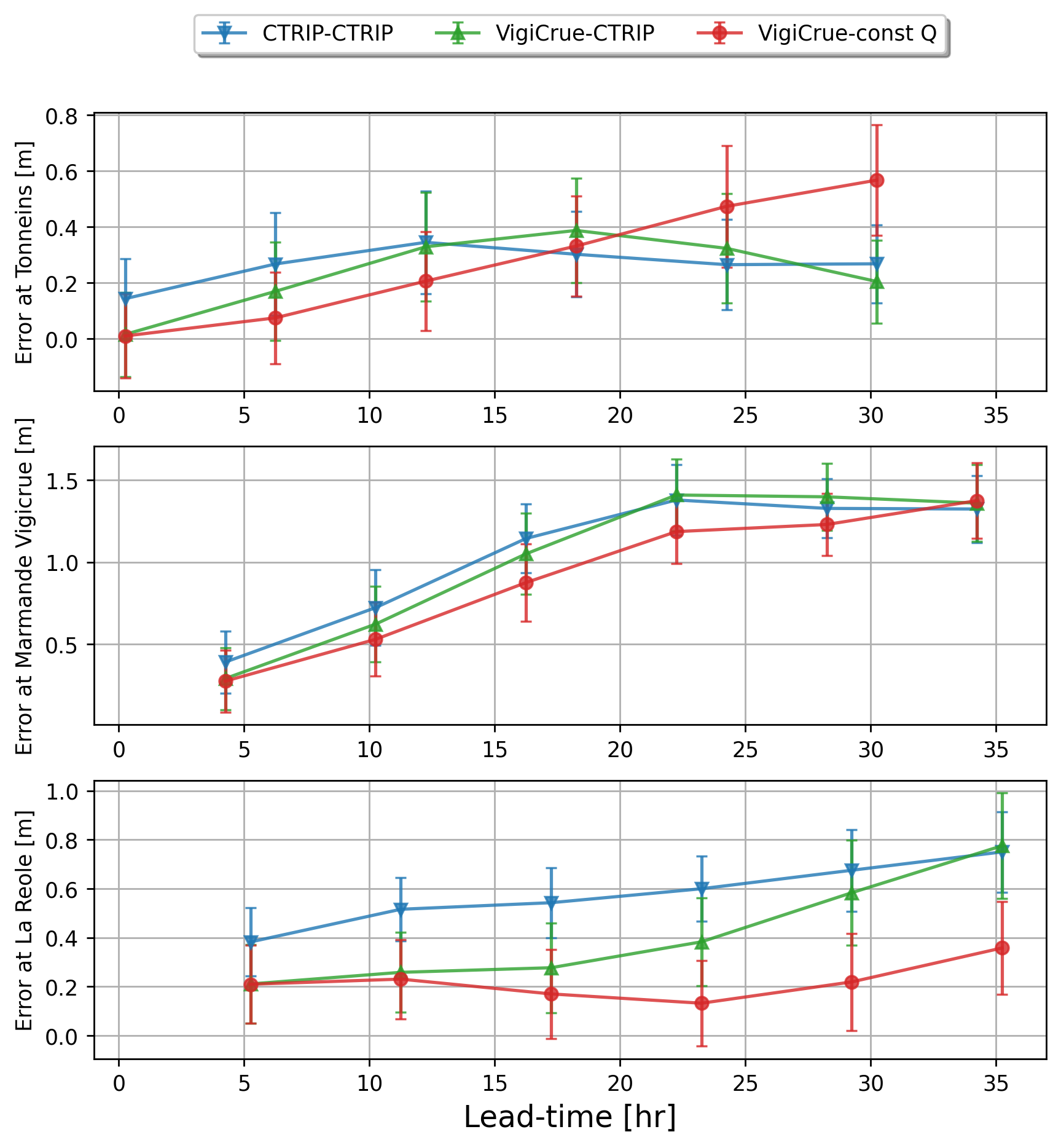}
    \caption{Forecast water level errors at Tonneins (target date 03/02/2021 12:15), Marmande (target date 03/02/2021 22:15), and La Réole (target date 04/02/2021 11:15) with various lead times. \fct}
    \label{fig:fct_error_targetdates}
\end{figure}


The forecasting capability of the DA strategy is evaluated over six days around the flood peak during the 2021 event at Tonneins, Marmande, and La Réole, considering increasing lead times. The forecasted water levels compared to the observations at Marmande are depicted in Figure~\ref{fig:DA2_fct_leadtime}. In the top panels, solid colored lines represent the forecast water levels with increasing lead time (+0~hr in blue, +6~hr in orange, +12~hr in green, +18~hr in red, +24~hr in purple, +30~hr in brown, and +36~hr in magenta), while the observed water levels are indicated by a black-dashed line. The corresponding error between the forecast and observation is depicted in the bottom panels using the same color code. 

Results indicate that the forecast succeeds in representing the flood rising limb for all lead times for $\mathrm{IGDA}$ using CTRIP-CTRIP  (Figure~\ref{fig:DA2_fct_leadtime}-(a)) and $\mathrm{IGDA}$ using VigiCrue-CTRIP (Figure~\ref{fig:DA2_fct_leadtime}-(b)) when CTRIP runoff is used as forcing beyond the present time $T_0$. On the opposite, prescribing a constant discharge (Figure~\ref{fig:DA2_fct_leadtime}-(c)) past the present time fails to represent the rising limb for lead times beyond 12~hr. It should also be noted that for all three forcing strategies, the flood peak is delayed and underestimated, especially beyond 6~hr lead time. The water level error reaches 3.5~m at Marmande during the flood recess on 06/02/2021 (Figure~\ref{fig:DA2_fct_leadtime}-(c), for +36~h). In spite of the limitations on the flood peak forecast, the success in properly forecasting the flood rising limb is satisfactory in terms of flood monitoring and alert. Therefore, the forcing strategies using CTRIP-CTRIP and VigiCrue-CTRIP  should be favored over VigiCrue-constant Q.

Figure~\ref{fig:RMSE_fct} depicts the RMSEs between $\mathrm{IGDA}$ using CTRIP-CTRIP, $\mathrm{IGDA}$ using VigiCrue-CTRIP, $\mathrm{IGDA}$ using VigiCrue-constant Q forecast and the observation. They are computed across six days around the flood peak for each observing station (Tonneins with a blue line, Marmande with an orange line, and La Réole with a green line) for increasing lead times. For all forcing strategies, as expected, the forecast performance decreases as the lead time increases, especially at Tonneins and Marmande. 
The performance between the three stategies are quite similar for a lead time below 12~hr, albeit with IGDA using VigiCrue-constant Q (Figure~\ref{fig:RMSE_fct}-(c)) showing slightly higher RMSEs. 

For short-term forecasts (under 12~hr of lead time), utilizing VigiCrue during the reanalysis phase proves most effective, followed by employing either CTRIP or constant Q during the forecast phase (between Figure~\ref{fig:RMSE_fct}-(b) and -(c)), which yield comparable results.
On the other hand, for longer lead times, employing either CTRIP or VigiCrue during the reanalysis phase results in similar RMSEs. However, CTRIP should be preferred over constant Q during the forecast phase. Ultimately, the most accurate forecast for longer lead times is achieved with $\mathrm{IGDA}$ using VigiCrue-CTRIP.


The forecast issued from the DA experiments are assessed here, focusing on the ensemble forecast mean and individual members. 
Errors in forecasted water levels at observing stations for the target date 01/02/2021 06:00, during the rising limb of the flood, are displayed in Figure~\ref{fig:fct_error_targetdate_20210201} (represented by the first vertical line in Figure~\ref{fig:DA2_fct_leadtime}).
The difference between the observed and the forecasted water levels is plotted for Tonneins (top panel), Marmande (middle panel) and La Réole (bottom panel) as a function of lead time varying from +0~hr to +36~hr along $x$-axis. The three forcing strategies---IGDA using CTRIP-CTRIP, using VigiCrue-CTRIP, and using VigiCrue-constant Q---are depicted in blue, green, and red, respectively.
This diagnostics involves seven DA cycles that provide an updated forecast every 6~hr for the target date, i.e. from 36~hr down to 0~hr prior to the target date.
At this stage, the flood event transitioned from a yellow to an orange alert level where overflows start to occur. It reemphasizes that $\mathrm{IGDA}$ using VigiCrue-CTRIP is the best strategy as previously concluded from Figure~\ref{fig:RMSE_fct}.

The results at the respective flood peak in forecast mode at the three observing stations are shown in Figure~\ref{fig:fct_error_targetdates}. They are depicted for the target dates 03/02/2021 12:15, 03/02/2021 22:15 (represented by the second vertical line in Figure~\ref{fig:DA2_fct_leadtime}), 04/02/2021 11:15, for flood peak at Tonneins, Marmande and La Réole, respectively. 
For all three forcing strategies, the forecast quality declines with increasing lead time, although there are exceptions. Notably, for Tonneins, both IGDA using CTRIP-CTRIP and using VigiCrue-CTRIP exhibit a slightly reduced error at lead times of 18~hr and beyond, due to the earlier peak discharge of CTRIP (see Figure~\ref{fig_4}). 
The most significant degradation in forecast is observed at Marmande, with overestimation that  exceeds 1~m  and reaches up to 1.5~m for lead times beyond 18~hr. 
Interestingly, when focusing on the flood peak target date (Figure~\ref{fig:fct_error_targetdates}), the third approach, i.e. IGDA using VigiCrue-constant Q, exhibits the lowest errors, particularly at Marmande and La Réole. 

\begin{figure*}[!t]
\centering
\includegraphics[width=\textwidth]{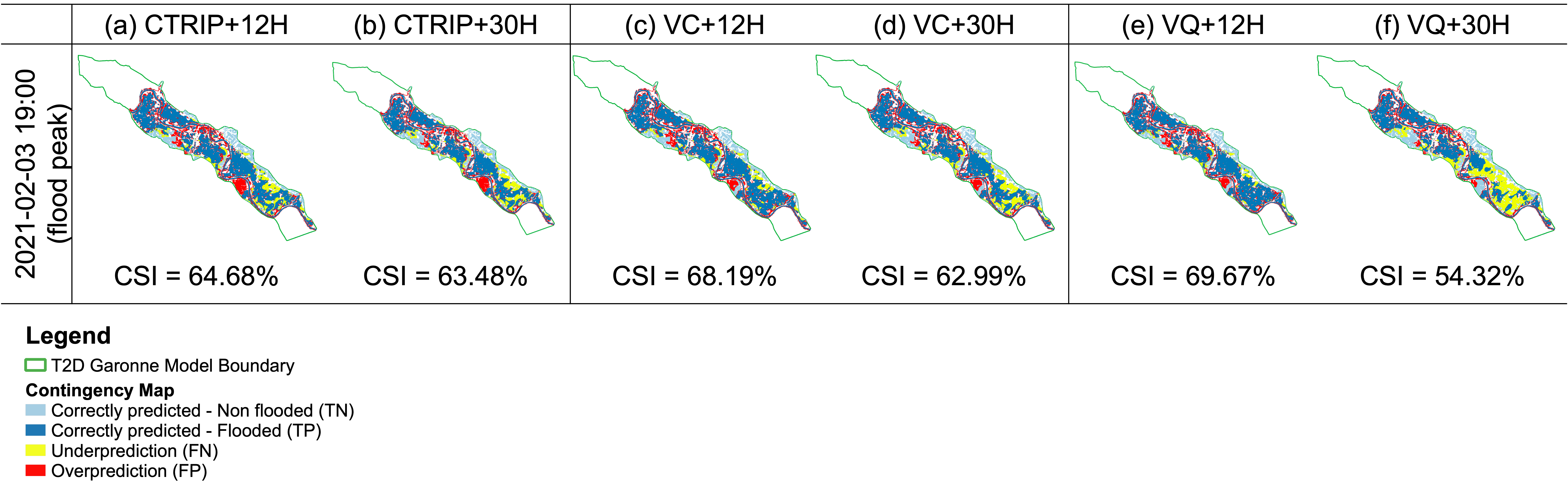}
\caption{Contingency maps and CSI computed between forecasted flood extents (averaged among $N_e$ members) and S1-derived observed flood extents for IGDA using the three different approaches at the rising limb and flood peak for the 2021 flood event. Column (a) and (b): IGDA using CTRIP-CTRIP with lead time +12~hr and +30~hr; Column (c) and (d): IGDA using VigiCrue-CTRIP with lead time +12~hr and +30~hr;  Column (e) and (f): IGDA using VigiCrue-constant Q with lead time +12~hr and +30~hr. \fct}
\label{fig:contingency_fct}
\end{figure*}

Figure~\ref{fig:contingency_fct} shows the contingency maps between the
forecasted flood extents and the observed flood extent derived from Sentinel-1 image.
Even with a 12-hour lead time, the forecast method employing IGDA with CTRIP-CTRIP nearly matches the renalysis CSI score of $\mathrm{IGDA}^\mathrm{C}$ previously shown in Figure~\ref{fig_8} (64.68\% compared to 68.34\%). Similarly, there is only a slight disparity between IGDA using VigiCrue-CTRIP and $\mathrm{IGDA}^\mathrm{V}$ (68.19\% versus 68.96\%). As expected, longer lead times result in lower CSI scores. However, a notable degradation between +12~hr and +30~hr (69.67\% and 54.32\%), occurs in the third approach, i.e. IGDA using VigiCrue-constant Q. This reemphasizes the large underestimations of the discharge for the VigiCrue-constant Q strategy (in the part leading up to the flood peak and the peak itself) when increasing lead times with this approach, as previously shown in Figure~\ref{fig:DA2_fct_leadtime}-(c).

\subsection{Post-event measure validation with High Water Marks observations}

The validation of the results in this work is also performed using independent observations from relevant high water marks (HWM) dataset. It is a collaborative dataset of collected flood marks\footnote{\url{https://www.reperesdecrues.developpement-durable.gouv.fr/}} maintained by the local flood forecast services (SPCs) or several \textquotedblleft \'{E}tablissements Publics Territoriaux de Bassins\textquotedblright~(EPTB) \cite{doi:10.1080/27678490.2024.2333410}. In the aftermath of the 2021 flood event, 178 HWM observations were collected (accessed in December 2022). 


\begin{table}[!t]
\caption{RMSE of simulated/forecasted maximum water levels and HWM observations. \reana \fct}
\label{tab:rmse_HWM}
\centering
\begin{tabular}{|c||c||c||c|}
\hline
& \multicolumn{3}{c|}{RMSE [m]}\\\hline
\multirow{4}{*}{Reanalysis} & $\mathrm{OL}^\mathrm{V}$ & $\mathrm{IDA}^\mathrm{V}$  & $\mathrm{IGDA}^\mathrm{V}$ \\\cline{2-4}
& 1.237 &  1.047 & 1.064 \\\cline{2-4}
& & & \\\cline{2-4}
& $\mathrm{OL}^\mathrm{C}$ & $\mathrm{IDA}^\mathrm{C}$  & $\mathrm{IGDA}^\mathrm{C}$ \\\cline{2-4}
& 1.695 & 1.077 & 1.068 \\\cline{2-4}
& & & \\\hline
\multirow{2}{*}{Forecast} & IGDA CC & IGDA VC & IGDA VQ \\
& +12~hr & +12~hr & +12~hr \\\cline{2-4}
& 1.171 & 1.146 & 1.085 \\\hline
\end{tabular}
\end{table}


A validation of the performed DA strategies and the three forecast approaches (focusing on IGDA) with respect to independent data was finally carried out using the collective public datasets of HWM for the 2021 flood event. This allows the evaluation of the highest simulated WLs spatially distributed at various points on the river banks and within the floodplain at the flood peak. 
Table~\ref{tab:rmse_HWM} summarizes the RMSE of the performed DA experiments and forecast approaches, between the simulated/forecast maximum water levels and the 178 HWM observations. 
In reanalysis, DA decreases the RSME to independent HWM observations when either VigiCrue or CTRIP forcing is used. In forecast mode (+12~h), the IGDA score remains close to that of the reanalysis (+0~h), meaning that the real-time forecasting workflow is beneficial. At this lead time, all three forcing strategies provide similar results. It is expected that IGDA with VigiCrue-CTRIP strategy would provide the best results for longer lead times. 



\section{Conclusion and Perspectives}
\label{concpersp}
This study presents the merits of assimilating 2D flood extent observations derived from Sentinel-1 SAR images with an EnKF implemented on the 2D hydraulic model TELEMAC-2D, focusing on a major flood event in 2021 over a catchment of the Garonne near Marmande. The control vector gathers friction coefficient and forcing correction and is augmented with correction of the hydraulic state in subdomains of the floodplains.  
The RS-derived flood extent observations are expressed in terms of WSR computed over defined sensitive subdomains of the floodplain. 

The DA framework is cycled with a real-time forecasting configuration. A cycle consists of a reanalysis phase and a forecast phase. Over the 6-hr reanalysis, observations up to the present are assimilated. An ensemble is then initialized from the last reanalyzed states and issued forecasts for the next 36~hr. 
In the reanalysis, six hindcast experiments were carried out both in Open Loop and Data Assimilation mode, using upstream forcing data that are either observed discharges from an observing station in the VigiCrue network, or discharges simulated by a large-scale hydrologic model ISBA-CTRIP. 
In the real-time forecast mode, three strategies of forcing data for this forecast are performed following IGDA reanalysis: (i) using CTRIP predicted runoff for analysis and forecast, (ii) using observed discharge for analysis and then CTRIP runoff for forecast and (iii) using observed discharge for analysis and keep a persistent discharge value for forecast. 

In hindcast mode, it was shown that IGDA greatly improves the water level in the riverbed and the representation of flow dynamics in the floodplain.
This improvement occurs when either observed discharge data from VigiCrue or large-scale hydrologic runoff from ISBA-CTRIP are utilized. 
The effectiveness of DA was assessed using both 1D and 2D evaluation metrics. 
While the assimilation of in-situ data allows to correct the hydraulic state in the riverbed, the assimilation of WSR further refines the flow dynamics in the floodplain, especially at flood peak and during recess. 
Indeed, IGDA  succeeds in identifying corrections to the friction coefficients and the hydrologic forcings so that the simulated hydraulic state is coherent with the assimilated in-situ and RS observations. 
This underscores the benefits of assimilating both in-situ and remote-sensing data for reanalyzing past flood events, leading to a more precise hydraulic state for initiating an ensemble of forecasts.

In forecast mode, the combination of observed forcing for the reanalysis phase and CTRIP-predicted forcing for the forecast phase provides the most accurate water levels and flood extent results. For all three investigated strategies, forecast quality diminishes with increasing lead time, particularly beyond 18~hr. 
Employing observed discharge or CTRIP runoff as forcing during the reanalysis phase yields comparable performance when CTRIP runoffs are used during the forecast phase. This approach proves superior to using a persistent Q value for forecast forcing, even though the quality of the CTRIP predicted runoff (in terms of amplitude and phase) may influence these conclusions at specific dates over a flood event. 


This work demonstrates that while imperfect, forcing data provided by a large-scale hydrologic model can be efficiently used as input to a local hydraulic model with DA in the context of reanalysis and real-time forecasting. These findings advocate for a multi-source strategy for the assimilation algorithm implemented on top of a chained hydrologic-hydraulic model that is favorable for short term as well as for extended lead time forecasts, especially in poorly gauged catchments.

\section*{Acknowledgments}
The authors gratefully thank Electricité de France (EDF) for providing the TELEMAC-2D model on the Garonne Downstream catchment, and the SCHAPI, SPCs Garonne-Tarn-Lot and Gironde-Adour-Dordogne for providing in-situ data. They would like to specially thank C. Fatras (CLS) for his developments of the FloodML algorithms, within the SCO-FloodDAM project, which are used in this research work. They also would like to thank E. Simon (IRIT) for fruitful discussions and advice concerning the Gaussian anamorphosis. Lastly, the authors would thank C. H. David and K. Marlis (NASA JPL) for their contributions in the original IGARSS 2023 paper regarding the use of RAPID model.

\bibliographystyle{IEEEtran}
\bibliography{IEEEabrv,ref}

\vfill

\end{document}